\documentclass[prd,a4paper,twocolumn,preprintnumbers,superscriptaddress,nofootinbib]{revtex4}
\pdfoutput=1
\usepackage{amsmath}
\usepackage{graphicx}
\usepackage{xspace}
\usepackage{color}
\usepackage{slashed}
\usepackage{units}
\usepackage[hyperfootnotes=false]{hyperref}
\usepackage[dvipsnames]{xcolor}

\def \L {\mathcal{L}} %Lagrangian density

\def \vec#1{{\boldsymbol{#1}}}
\newcommand{\matrixx}[1]{\begin{pmatrix} #1 \end{pmatrix}} %Matrix with brackets
\newcommand{\hc}{\ensuremath{\text{h.c.}}}
\newcommand{\dd}{\mathrm{d}}
\newcommand{\tr}{\mathrm{tr}}

\allowdisplaybreaks

\makeatletter
\newcommand\footnoteref[1]{\protected@xdef\@thefnmark{\ref{#1}}\@footnotemark}
\makeatother

\begin{document}
%%%%%%%%%%%%%%%%%%%%%%%%%%%%%%%%%

\title{Cold keV dark matter from decays and scatterings}

\preprint{ULB-TH/17-09}

\author{Julian Heeck}
\email{Julian.Heeck@ulb.ac.be}
\affiliation{Service de Physique Th\'eorique, Universit\'e Libre de Bruxelles, Boulevard du Triomphe, CP225, 1050 Brussels, Belgium}

\author{Daniele Teresi}
\email{Daniele.Teresi@ulb.ac.be}
\affiliation{Service de Physique Th\'eorique, Universit\'e Libre de Bruxelles, Boulevard du Triomphe, CP225, 1050 Brussels, Belgium}

\hypersetup{
    pdftitle={
		Cold keV dark matter from decays and scatterings},
    pdfauthor={Julian Heeck, Daniele Teresi}
}

%%%%%%%%%%%%%%%%%%%%%%%%%%%%%%%%%%
%%%%%%%%%%%%%%%%%%%%%%%%%%%%%%%%%%

\begin{abstract}

We explore ways of creating cold keV-scale dark matter by means of decays and scatterings. The main observation is that certain thermal freeze-in processes can lead to a cold dark matter distribution in regions with small available phase space. In this way the free-streaming length of keV particles can be suppressed without decoupling them too much from the Standard Model. In all cases, dark matter needs to be produced together with a heavy particle that carries away most of the initial momentum. For decays, this simply requires an off-diagonal DM coupling to two heavy particles; for scatterings, the coupling of soft DM to two heavy particles needs to be diagonal, in particular in spin space. Decays can thus lead to cold light DM of any spin, while scatterings only work for bosons with specific couplings. We explore a number of simple models and also comment on the connection to the tentative 3.5 keV line.

\end{abstract}

%%%%%%%%%%%%%%%%%%%%%%%%%%%%%%%%%
%%%%%%%%%%%%%%%%%%%%%%%%%%%%%%%%%
\maketitle

%%%%%%%%%%%
\tableofcontents

\section{Introduction}

Dark matter (DM) has been indirectly observed via its gravitational interactions, be it in structure formation, galaxy rotation curves or lensing. Additional couplings of DM to the particles of the Standard Model (SM) have been searched for intensely, but so far without an undisputed discovery. Consequently, we have little information about the properties of DM beyond its average abundance, apparent stability, and ability to form structures. The latter implies that DM should be sufficiently \emph{cold}, i.e.~non-relativistic, in order to not wash out the observed small structures in the Universe. For DM particles motivated by the so-called WIMP miracle (Weakly Interacting Massive Particles) this happens quite naturally for DM masses above GeV. Persistent negative results in both direct and indirect detection are, however, putting pressure on the standard WIMP paradigm and have led to an increase in alternative DM scenarios.

One popular class here are models with DM particles around the keV scale. For fermion DM, keV corresponds to the smallest mass that still allows to form the small structures that we observe in our Universe; this Tremaine--Gunn bound~\cite{Tremaine:1979we,Boyarsky:2008ju} follows from Fermi--Dirac statistics and holds independently of the DM production mechanism.
No such strict lower bound exists for bosonic DM, with many models going far below the keV scale, most prominently discussed for axion DM~\cite{Olive:2016xmw}.
Still, if keV-scale DM of any spin is produced \emph{thermally} in the broadest sense, one has to worry about the formation of small structures. Particles of this kind are then candidates for \emph{warm} DM, meaning they possess a sizable free-streaming length $\lambda_\mathrm{fs}$ of the order of \unit[0.1]{Mpc} that washes out structures below $\lambda_\mathrm{fs}$ but behave similar to cold DM on larger scales. This might be a better description of our Universe than cold DM, which possibly predicts too many small structures, judging by state-of-the-art simulations. Reconcilement with observations might lie in the inclusion of baryons in the simulations, which is an ongoing effort in several groups~\cite{Fattahi:2016nld}.
While $N$-body simulations might hint at an $\mathcal{O}(\unit[0.1]{Mpc})$ free-streaming length, there are competing constraints on structures of this size from Lyman-$\alpha$ data~\cite{Yeche:2017upn}; the inclusion of higher-resolution data from HIRES and MIKE~\cite{Viel:2013apy} leads to particularly strong bounds, but have to be interpreted with care in view of potential astrophysical effects~\cite{Garzilli:2015iwa,Baur:2017stq}.
More simulations and data are necessary to settle this point, but for now it seems that DM free-streaming lengths slightly below the Lyman-$\alpha$ bounds could still play a role in solving some of the structure-formation issues that cold DM faces.

Besides structure formation, keV-scale DM also offers interesting indirect detection signatures. Since only photons and neutrinos are possible final states for a decay, it typically does not take absurdly small couplings to make keV particles stable on cosmological scales. Compared to WIMPs, it is therefore often not necessary to impose an artificial stabilizing symmetry. X-ray photons from such decaying DM can then be searched for in astrophysical objects with large DM density, and feature a different morphology than photons from annihilating DM.
Interestingly, a line-like signal at the photon energy $E_\gamma \simeq \unit[3.55]{keV}$ was observed in 2014 independently by two different groups~\cite{Bulbul:2014sua,Boyarsky:2014ska}. While the significance and mere existence of this line is heavily debated, there is no clear consensus as of now in the community. We will consider this tantalizing excess as an interesting benchmark point in this article, but stress that our work is applicable in a more general context.
Taking the \unit[3.55]{keV} line seriously in the context of two-body DM decay, updated Lyman-$\alpha$ constraints disfavor or even exclude almost all currently proposed production mechanisms for $\unit[7]{keV}$ DM~\cite{Merle:2014xpa,Schneider:2016uqi}, providing ample motivation for the ideas discussed in the following. 

KeV-scale \emph{fermion} DM is most prominently discussed in the context of sterile-neutrino DM -- exhaustively reviewed in Refs.~\cite{Boyarsky:2009ix,Kusenko:2009up,Konig:2016dzg,Adhikari:2016bei,Abazajian:2017tcc} -- while \emph{bosonic} DM is often envisioned as the pseudo-Goldstone boson of an approximate global symmetry; in this way, huge radiative corrections to the scalar DM mass can be avoided. Examples here are majorons~\cite{Rothstein:1992rh,Berezinsky:1993fm,Lattanzi:2007ux,Bazzocchi:2008fh,Frigerio:2011in,Lattanzi:2013uza,Queiroz:2014yna}, connected to the lepton symmetry $U(1)_L$~\cite{Chikashige:1980ui,Schechter:1981cv}, familons~\cite{Frigerio:2011in}, connected to family symmetries~\cite{Wilczek:1982rv,Reiss:1982sq}, and axions (or axion-like) particles~\cite{Higaki:2014zua,Jaeckel:2014qea}, connected with the Peccei--Quinn symmetry $U(1)_\text{PQ}$~\cite{Kim:1986ax}.
In order to not violate Lyman-$\alpha$ constraints, the production mechanism of any keV-scale DM is required to be always out of thermal equilibrium. The constraints are in fact becoming so strong that even many out-of-equilibrium mechanisms are already excluded, forcing the DM to be increasingly weakly coupled, e.g.~by successive freeze-in mechanisms~\cite{Merle:2013wta,Biswas:2016bfo,Biswas:2016iyh}.

In this article we are going to present new ways to produce keV-scale DM from thermal processes without violating Lyman-$\alpha$ bounds by making use of kinematical features in the way discussed at length below. A key aspect of our mechanisms is that the parent particle(s) from which DM is produced by decays or scattering can be at thermal equilibrium with the SM during DM production. This generically makes DM more testable than other existing mechanisms (see however Refs.~\cite{Molinaro:2014lfa,Hessler:2016kwm}) to produce cold enough keV DM based, for instance, on a double freeze-in of both DM and the parent particle. While the latter typically require very small couplings of both DM and the parent particle to the SM, the mechanisms discussed below allow (or even \emph{require}, in some explicit models) large couplings of the parent particle(s) with the SM, thus making this testable at current or future experiments.

The rest of the article is structured as follows: in Sec.~\ref{sec:lyman} we discuss Lyman-$\alpha$ limits, which give the most important constraints on the keV-scale DM scenarios under consideration here. Sec.~\ref{sec:technicalities} lists the Boltzmann equations necessary for our discussion and provides useful formulae for DM observables. Our main ideas for cold light DM are illustrated in Sec.~\ref{sec:toy_models} with some simple toy models, while more realistic scenarios are discussed in Sec.~\ref{sec:full_models}. In Sec.~\ref{sec:decay} we finally discuss DM decays in connection with the \unit[3.5]{keV} line, and conclude our work in Sec.~\ref{sec:conclusion}.
Appendix~\ref{sec:formulae} gives technical details about the Boltzmann equations we employ in the main text, while App.~\ref{sec:xsecs} lists some of the cross sections relevant for our discussion.

\section{Small-scale structure constraints}
\label{sec:lyman}

KeV-scale DM has been discussed extensively in the literature and is often motivated as a warm DM candidate.
It must be stressed, however, that keV DM particles do not automatically form \emph{warm} DM, because the ability to form structures depends not on the mass but on the actual momentum distribution $f(p,T)$ of the particle. This, in turn, depends strongly on the DM production mechanism and the details of the following cosmological evolution. 
Since $f(p,T)$ can be arbitrarily complicated, it is non-trivial to constrain its form from astrophysical data~\cite{Murgia:2017lwo}. As an approximation, it is often sufficient to define a simple free-streaming length that contains the momentum-average of $f(p,T)$. Limits from small-scale structure observations are typically given for two \emph{benchmark} scenarios of keV-scale fermion DM, which subsequently need to be translated to bosonic DM and more general production mechanisms of interest to us.
\begin{enumerate}
	\item \emph{Thermal Relic} (TR) DM, which was in thermal equilibrium down to a decoupling temperature $T_D\gg \unit{keV}$. The momentum distribution is then simply Fermi--Dirac (FD), but with an entropy-diluted temperature 
	\begin{align}
	T_\mathrm{DM}= \left(\frac{g_*(T)}{g_*(T_D)}\right)^{1/3} T \,.
	\end{align}
	This momentum distribution gives a free-streaming length~\cite{Colombi:1995ze}
	\begin{align}
	\lambda_\mathrm{fs}\sim \unit{Mpc} \left(\frac{\unit{keV}}{m_\mathrm{TR}}\right) \left(\frac{g_*(T_\nu)}{g_*(T_D)}\right)^{1/3}
	\end{align}
	and DM abundance 
	\begin{align}
	\Omega_\mathrm{DM}h^2 \simeq \left(\frac{g_*(T_\nu) }{g_*(T_D)}\right) \left(\frac{m_\mathrm{TR}}{\unit[94]{eV}}\right),
	\end{align}
	where $g_*(T_\nu) = 10.75$ is the entropy number of degrees of freedom before neutrino decoupling, which arises when translating the well-known neutrino abundance~\cite{Bond:1980ha} to a different species.
	The decoupling temperature $T_D$ can be eliminated from $\lambda_\mathrm{fs}$ using $\Omega_\mathrm{DM}$, which results in a characteristic scaling $\lambda_\mathrm{fs}\propto m_\mathrm{TR}^{-4/3}$. 
	For $\Omega_\mathrm{DM}h^2 \simeq 0.12$, Lyman-$\alpha$ analyses (including high-resolution data~\cite{Viel:2013apy}) give a constraint of $m_\mathrm{TR}\gtrsim \unit[4.65]{keV}$ on this scenario~\cite{Yeche:2017upn,Baur:2017stq}, which would in turn require an entropy factor $g_*(T_D)\gtrsim 4400$ to obtain the correct relic abundance. Seeing as the SM can only provide for $g_* (T>\unit[200]{GeV}) = 106.75$, this would necessitate an enormously large number of new particles at higher temperatures (or another form of entropy release after DM decoupling~\cite{Bezrukov:2009th,Babu:2014uoa}). While not impossible, this scenario is clearly disfavoured, thus forcing keV DM particles to be out of equilibrium.
Note that Ref.~\cite{Irsic:2017ixq} gives an even stronger bound, $m_\mathrm{TR}\gtrsim \unit[5.3]{keV}$, but we will not use it here to stay somewhat conservative.
The impact of warm DM on the ionization history of our Universe was recently studied in Ref.~\cite{Lopez-Honorez:2017csg}, but does not yet provide competitive constraints on $m_\mathrm{TR}$.

\item \emph{Non-Resonant Production} (NRP)~\cite{Dodelson:1993je}, where the sterile neutrino is produced out-of-equilibrium at temperatures $\unit[\mathcal{O}(100)]{MeV}$ from oscillations of active neutrinos with a FD distribution $f(p,\left(4/11\right)^\frac13 T)$ that is suppressed by the active--sterile mixing angle~\cite{Dodelson:1993je,Colombi:1995ze}. The relic abundance is then effectively a free parameter, and the free-streaming length is simply
	\begin{align}
	\lambda_\mathrm{fs}\sim \unit{Mpc} \left(\frac{\unit{keV}}{m_\mathrm{NRP}}\right) \,.
	\label{eq:NRP}
	\end{align}
	Current limits on this scenario are  $m_\mathrm{NRP}\gtrsim\unit[28.8]{keV}$~\cite{Yeche:2017upn}.
Fixing the relic abundance $\Omega_\mathrm{DM}h^2\simeq 0.12$ allows one to derive a relation between $m_\mathrm{NRP}$ and $m_\mathrm{TR}$ that leads to the same free-streaming length~\cite{Colombi:1995ze},
\begin{align}\label{eq:TRtoNRP}
m_\mathrm{NRP}\simeq \unit[4.5]{keV}\left(\frac{m_\mathrm{TR}}{\unit{keV}}\right)^{4/3} \,.
\end{align}
Together with $m_\mathrm{TR}\gtrsim \unit[4.65]{keV}$ this gives a similar bound on $m_\mathrm{NRP}$ as a full analysis~\cite{Yeche:2017upn}. The limited discrepancy is due to the fact that~\eqref{eq:TRtoNRP} does not take into account the change in the number of degrees of freedom during the QCD phase-transition~\cite{Abazajian:2005gj,Bozek:2015bdo}.
Note that this mechanism, too, is disfavored when combined with x-ray limits~\cite{Abazajian:2017tcc}.
\end{enumerate}

The bounds on $m_\mathrm{TR}$ or $m_\mathrm{NRP}$ cannot, strictly speaking, be translated to models with different production mechanisms, for which a complete re-analysis with the full momentum distribution $f(p,T)$ is required. This is beyond the scope of this work, since we will explore a multitude of different production mechanisms and thus need a quicker way to asses their resulting phenomenology. As such, we will make the simplifying assumption that the free-streaming length of our keV-scale DM particles is similar to Eq.~\eqref{eq:NRP}, but weighted by the average DM momentum over temperature $\langle p/T\rangle$, which is given by its value at production time, times the usual entropy dilution factor, defined by the production or decoupling temperature $T_D$,
	\begin{align}
	\left\langle \frac{p}{T}\right\rangle = \left\langle \frac{p}{T}\right\rangle_\mathrm{prod} \left(\frac{g_*(T)}{g_*(T_D)}\right)^{1/3} .
	\label{eq:poverTtoday}
	\end{align}
	The limits on $m_\mathrm{NRP}$ can then be approximately translated to other production mechanisms using~\cite{Shaposhnikov:2006xi,Bezrukov:2014nza,Adhikari:2016bei}
	\begin{align}
 m_\mathrm{DM} = \frac{\langle p/T\rangle m_\mathrm{NRP}}{\langle p_\mathrm{NRP}/T\rangle }= \frac{\langle p/T\rangle_\mathrm{prod}}{3.15} \left(\frac{10.75}{g_*(T_D)}\right)^{\frac{1}{3}} m_\mathrm{NRP} \;.
\label{eq:translation}
	\end{align}
Notice that $m_{\rm NRP}$ in this formula should be as given by~\eqref{eq:TRtoNRP}, i.e.~without the modifications due to the QCD phase-transition mentioned above, since here $g_*$ at neutrino decoupling has been fixed to the value $10.75$.
A lower bound on the DM mass can then be obtained given the average DM momentum at production temperature $T_D$.
Alternatively, an upper bound on $\langle p/T\rangle_\mathrm{prod}$ can be derived for a given DM mass,
\begin{align}
\left\langle \frac{p}{T}\right\rangle_\mathrm{prod} \lesssim \frac{m_\mathrm{DM}}{\unit[5.1]{keV}} \left(\frac{g_*(T_D)}{106.75}\right)^{\frac{1}{3}} \left(\frac{\unit[4.65]{keV}}{m_\mathrm{TR}}\right)^{\frac{4}{3}} .
\label{eq:translation2}
\end{align}
This is a particularly convenient constraint on the DM production mechanism if the DM mass is fixed by other means, e.g.~from indirect detection. 
Assuming, for example, the DM to be produced above the electroweak phase transition (EWPT), a \unit[7]{keV} DM particle needs to be produced with an average momentum 
\begin{align}		
\langle p/T\rangle_\mathrm{prod}  \lesssim 1.4 \,,
\label{eq:7kevLyman}
\end{align}		
to satisfy the Lyman-$\alpha$ bound $m_{\rm TR} \gtrsim \unit[4.65]{keV}$. This is far below the thermal FD value of  $\langle p/T\rangle_\mathrm{thermal}^\mathrm{FD}\simeq 3.15$ or the value $2.45$ one obtains by DM production via decays of heavy thermalized scalars $S\to \mathrm{DM}\,\mathrm{DM}$~\cite{Shaposhnikov:2006xi,Bezrukov:2014nza}.
See Refs.~\cite{Kusenko:2009up,Shakya:2015xnx,Konig:2016dzg,Adhikari:2016bei,Abazajian:2017tcc} for other models under consideration in the literature.
Below, we will explore various production mechanisms that can indeed yield such low momenta and are thus capable of accommodating much lighter DM than in other models.

An important remark is at order:
Eq.~\eqref{eq:translation} is accurate only if the momentum distribution has a meaningful mean value $\langle p/T \rangle$ to begin with. If $f(p,T)$ looks highly non-thermal, e.g.~with two components peaked at momenta that differ by one or more orders of magnitude, the proper way to estimate Lyman-$\alpha$ constraints is to go back to the transfer function and evaluate everything numerically~\cite{Bezrukov:2014qda,Konig:2016dzg} or even to consider directly structure-formation simulations with non-thermal spectra~\cite{Murgia:2017lwo}. We will come back to this issue in future work; for the purposes of this article it will be sufficient to work at the level of average momentum $\langle p/T \rangle$.

\section{Boltzmann equations}
\label{sec:technicalities}

In this section, we will discuss how to obtain the momentum distribution $f$ of light DM produced by decays and/or scatterings.

In the cases of interest here, all but the DM particle are thermalized, so that the system can be described by just the Boltzmann equation for the DM. This corresponds to the \emph{freeze-in} scenario, where we assume a negligible initial DM abundance~\cite{McDonald:2001vt,Hall:2009bx,Bernal:2017kxu}. In order to obtain Boltzmann equations in a simple form, for instance expressed in terms of scattering cross sections, we will approximate the distribution functions as given by classical statistics, i.e.~Maxwell--Boltzmann (MB) distributions. Thus, for a thermalized particle the thermal average for the momentum is $\langle p/T\rangle_\mathrm{thermal}^\mathrm{MB} = 3$. Also, notice that the very form of the Boltzmann equations below would be modified in the presence of a sizeable number density of particles in the final state if quantum statistics were to be used. As we are going to show later on, the usage of classical statistics introduces an uncertainty of a few percent on the momentum averages obtained below. We will also neglect the change in the number of relativistic degrees of freedom $g_*$ during DM production. These issues may be addressed elsewhere in a more elaborate numerical study, together with the other uncertainties present in our calculation. However, these are not expected to affect the conclusions of the analysis given here. Further details about the derivation of these Boltzmann equations are given in Appendix~\ref{sec:formulae}.

\subsection{Scatterings}

We consider the $2\to 2$ scattering of two thermalized particles $A B \to C\,  \mathrm{DM}$, where the particles involved can be bosons or fermions. Since the \unit{keV}-scale DM candidate is produced by freeze-in, the Boltzmann equation depends only on the distribution functions of the (thermalized) particles, which we take to be MB distributions in the approximation of classical statistics. We denote by $m_X$ the mass of $X \in \{ A,B, C\}$, which we envision to be around or above the electroweak scale. The DM particle on the other hand is approximated as massless in comparison, having in mind \unit{keV}-scale DM. We introduce the dimensionless variables 
\begin{align}
r \equiv \frac{m_H}{T}\,, \ \  x \equiv \frac{|\vec{p}|}{T} \,,  \ \  a_X \equiv\frac{m_X^2}{m_H^2} \,, \  X \in \{ A,B, C\}\,,
\end{align}
where $\vec{p}$ is the 3-momentum of the DM particle with magnitude $|\vec{p}|\equiv p$. For bookkeeping purposes we have introduced the reference mass scale $m_H$, the Higgs-boson mass, which will of course drop out of physical quantities. We find the Boltzmann equation for the DM distribution function $f(x,r)$ relevant for scattering processes $A B \to C\,  \mathrm{DM}$ as
\begin{align}\label{eq:eq_scat}
\frac{\partial f}{\partial r}  &= \frac{1}{16 \pi^2}\frac{M_0}{m_H}  \, \frac{1}{x^2} \int_{y^*}^{\infty} \!\! \dd y \; \widehat{\sigma}\Big(\frac{m_H^2 y}{r^2}\Big) \; \bigg(1 - \frac{a_C r^2}{y} \bigg)^{-1} \notag\\
&\quad\times \exp\left[- x \, \bigg(1 - \frac{a_C r^2}{y} \bigg)^{-1} \!\! - \frac{y - a_C r^2}{4 x}\right]
\end{align}
where $M_0 \equiv M_\text{Pl} \sqrt{45/(4\pi^3 g_*)}$. Here, $\widehat{\sigma}(s)$ is the reduced cross section~\cite{Luty:1992un} for the process $A B \leftrightarrow C\, \mathrm{DM}$,\footnote{In obtaining Eq.~\eqref{eq:eq_scat} we have neglected possible CP violation in the process, so that the reduced cross section is T-invariant and does not depend on the direction of the reaction.} summed over the initial- and final-state's internal degrees of freedom, and
\begin{align}
y^* \equiv r^2\times \max \big\{(\sqrt{a_A} + \sqrt{a_B})^2, \, a_C \big\} 
\end{align}
defines the kinematic threshold for the process.
An analytic integration of Eq.~\eqref{eq:eq_scat} is not possible for a general cross section, so let us simply state how to calculate the relevant quantities once $f(x,r)$ has been obtained, for example numerically. The DM abundance is obtained by integrating the distribution function over momentum, i.e.~$\int \dd^3  \vec p \, f(\vec p,\infty)$, and normalizing to the entropy density in the standard way~\cite{Kolb:1990vq},
\begin{align}
\Omega_\text{DM}h^2 = \frac{s_0 m_\mathrm{DM}}{\rho_\mathrm{crit}/h^2}\left[\frac{45/(4\pi^4)}{g_*(T_\mathrm{prod})} \int_0^\infty \dd x \, x^2 f(x,\infty) \right],
\label{eq:abundance}
\end{align}
where the quantity in brackets is the yield and the DM distribution is evaluated at $r\to \infty$, i.e.~today. For a definition of the entropy density $s_0$ and critical density $\rho_\mathrm{crit}$ we refer to the standard literature, e.g.~Ref.~\cite{Kolb:1990vq}.
The average DM momentum at production is given by the first moment of the distribution function, i.e.
\begin{align}
 \left\langle \frac{p}{T}\right\rangle_\mathrm{prod} = \frac{\int \! \dd^3  \vec p \, |\vec p|  f(\vec p,r_{\rm prod})}{\int \! \dd^3  \vec p \, f(\vec p,r_{\rm prod})} = \frac{\int_0^\infty \dd x \, x^3 f(x,r_\mathrm{prod})}{\int_0^\infty \dd x \, x^2 f(x,r_\mathrm{prod})} \,.
\label{eq:mean_pT}
\end{align}
Since our calculations assume a constant $g_*$, we may take the limit $r_\mathrm{prod} \to \infty$; notice, however, that the average DM momentum \emph{today}, which is instead defined in Eq.~\eqref{eq:poverTtoday}, involves also the entropy dilution after DM production.

The DM couplings to the SM are assumed to be small by construction, but one could imagine DM being coupled more strongly to other particles, a full dark sector. Thermalization within that sector would then convert the DM distribution $f$ obtained via freeze in into a standard thermal distribution, albeit with a different temperature than the SM. The mean momentum of Eq.~\eqref{eq:mean_pT} then trivially reduces back to a thermal value. Discussions of such scenarios can be found in Refs.~\cite{Heikinheimo:2016yds,Heikinheimo:2017ofk,Hansen:2017rxr,Bernal:2017kxu}; we will not consider this further, but rather assume that the DM distribution is entirely determined by the freeze-in process.

\subsection{Decays}

Let us now consider the 2-body decay of a thermalized particle $A \to B \, \mathrm{DM}$ under the same assumptions and definitions as above. The DM particle is again approximated as massless, and in particular also much lighter than the mass \emph{difference} of $A$ and $B$, i.e.~$m_\mathrm{DM}\ll m_A-m_B \leq m_A$. We introduce the additional variable
\begin{align}
\Delta \equiv 1-\frac{m_B^2}{m_A^2} \,,
\end{align}
which is a measure of the phase-space closure in the decay and takes on values between 0 and 1. We find the evolution equation for the DM distribution $f(x,r)$ as
\begin{align}
\frac{\partial f}{\partial r}  &= \frac{g_A\,S\,\Gamma\,M_0\sqrt{a_A}}{m_H^2 \Delta}\frac{r^2}{x^2} \int\limits_{\frac{x}{\Delta}+\frac{r^2}{4}\frac{\Delta}{x}a_A}^\infty \dd\xi \,f_{A}\!\!\left(\tfrac{E}{T}=\xi\right) \label{eq:boltzmann_decay}\\
&\overset{\text{MB}}{\to}\frac{g_A\,S\,\Gamma\,M_0\sqrt{a_A}}{m_H^2 \Delta}\frac{r^2}{x^2} \,\exp\left(-\frac{x}{\Delta}-\frac{r^2}{4}\frac{\Delta}{x}a_A\right) ,\label{eq:boltzmann_decay_MB}
\end{align}
where $\Gamma$ is the partial decay width of the process $A \to B \, \mathrm{DM}$, $g_A$ counts the internal degrees of freedom of $A$, and $S$ is the symmetry factor of the decay, equal to 2 if $B = \mathrm{DM}$ and 1 otherwise.
Eq.~\eqref{eq:boltzmann_decay} generalizes the known result for the $S\to N N$ decay~\cite{Shaposhnikov:2006xi,Petraki:2007gq}.
In Eq.~\eqref{eq:boltzmann_decay_MB} we have used the MB form for the equilibrium distribution function of $A$, $f_{A}(E) = \exp (-E/T)$.

From Eq.~\eqref{eq:boltzmann_decay} we see that the DM momentum always appears in the combination $x/\Delta$, except for the overall normalization of $f$. As a result, the mean $\langle x\rangle$ of Eq.~\eqref{eq:mean_pT} always scales \emph{linearly} with the phase-space suppression factor~$\Delta$, independent of the distribution of the mother particle~$A$. The prefactor of this scaling, however, depends on $f_{A}$ and can be calculated as
\begin{align}
\left\langle \frac{p}{T}\right\rangle_\mathrm{prod}^\mathrm{MB} \,&= \, \frac{5}{2} \,\Delta \,=\, 2.5\Delta\,,\label{eq:pT_decay}\\
\left\langle \frac{p}{T}\right\rangle_\mathrm{prod}^\mathrm{BE} \,&= \,\frac{\pi^6}{378 \zeta (5)} \, \Delta \,\simeq \, 2.45\Delta\,,\label{eq:pT_decay_BE}\\
\left\langle \frac{p}{T}\right\rangle_\mathrm{prod}^\mathrm{FD} \, &= \, \frac{31\pi^6}{11340 \zeta (5)} \, \Delta \, \simeq\,  2.53\Delta \,,
\end{align}
for the three distribution functions of interest, MB, FD, and Bose--Einstein (BE). This linear scaling holds for $m_\mathrm{DM}\ll m_A-m_B$, but breaks down once the DM mass becomes non-negligible.

Eq.~\eqref{eq:boltzmann_decay_MB} can be easily integrated analytically for the MB distribution, leading in particular to the DM distribution function today,
\begin{align}
f(x,\infty) = \frac{2 \sqrt{\pi } g_A\,S\,\Gamma\,M_0}{m_A^2 \Delta ^3} \sqrt{\frac{\Delta}{x}} \exp\left(-\frac{x}{\Delta}\right) ,
\label{eq:decayf0}
\end{align}
which gives the DM abundance via Eq.~\eqref{eq:abundance},
\begin{align}
\Omega_\text{DM}h^2 &= \frac{135}{8\pi^3 g_*(T_\mathrm{prod})} \frac{s_0 m_\mathrm{DM}}{\rho_\mathrm{crit}/h^2}\frac{g_A\,S\, \Gamma\,M_0}{m_A^2} \label{eq:abundance_decay}\\
&\sim 0.08 \,g_A \left(\frac{m_\mathrm{DM}}{\unit[7]{keV}}\right)\left(\frac{\unit[1]{TeV}}{m_A}\right)\left(\frac{S \, \Gamma/m_A}{10^{-17}}\right) \nonumber .
\end{align}
Using instead the BE distribution for $f_A$ increases the abundance by $\zeta (5)\simeq 1.037$~\cite{Merle:2015oja}, whereas it is lowered by $\tfrac{15}{16}\zeta (5)\simeq 0.972$ in the FD case. 
Eq.~\ref{eq:abundance_decay} matches Refs.~\cite{Hall:2009bx,Merle:2015oja,Adulpravitchai:2015mna} in their cases of interest.
We stress that the \emph{total} lifetime of $A$ is not restricted to be long, as it is only the \emph{partial} width $A\to B\,\mathrm{DM}$ that enters Eq.~\eqref{eq:abundance_decay} and therefore has to be small.

Finally, let us mention that most of the DM \emph{abundance} of Eq.~\eqref{eq:abundance_decay} is created around $m_A/T \sim 3$--$5$~\cite{Hall:2009bx,Frigerio:2011in}, but it takes until $m_A/T \sim 10$ to reach the average-momentum plateau $\langle p/T\rangle\simeq 2.5 \Delta$, meaning that the hottest part of the spectrum is created the latest.

\subsection{Multiple production processes}\label{sec:multiple}

\begin{figure}[t]
\includegraphics[width=0.48\textwidth]{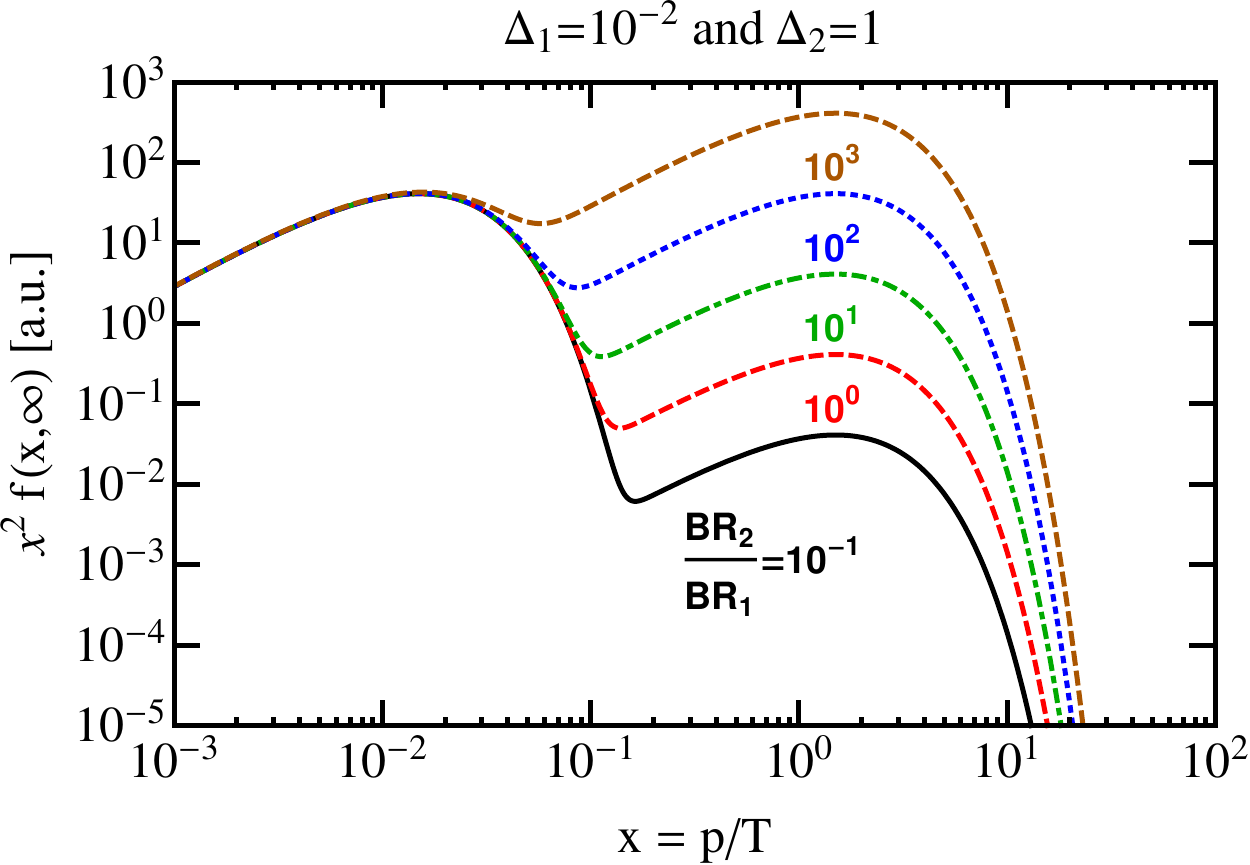}
\caption{
DM distribution $x^2 f(x, r\to \infty)$ in arbitrary units for the case of two competing decays $A\to B_j\,\mathrm{DM}$, $j=1,2$, with branching ratios $\mathrm{BR}_j$ and phase space suppression factors $\Delta_j = 1-m_{B_j}^2/m_A^2$.
}
\label{fig:two_decays}
\end{figure}

In the case of several (decay and/or scattering) processes, their effect on $f(x,r)$ is simply additive, since we are in the freeze-in regime for DM where the inverse processes depending on $f(x,r)$ do not give sizeable contributions.
This implies that the effect is also linear for the DM abundance $\Omega_\text{DM}h^2$ (Eq.~\eqref{eq:abundance}), whereas the impact on $\langle p/T\rangle$ (Eq.~\eqref{eq:mean_pT}) is more complicated and can even make the usefulness of an average momentum questionable.
As a simple example, let us consider two competing decays $A\to B_j\,\mathrm{DM}$, $j=1,2$, with branching ratios $\mathrm{BR}_j$ and phase space suppression factors $\Delta_j = 1-m_{B_j}^2/m_A^2$. For definiteness, let us assume $\Delta_1 \ll \Delta_2 \simeq 1$, so that $\mathrm{BR}_2/\mathrm{BR}_1$ determines the ratio of hot to cold DM. If there is a large hierarchy between $\mathrm{BR}_1$ and $\mathrm{BR}_2$, the mean $x$ is a sensible measure for the impact on small-scale structure; however, for 
\begin{align}
\frac{\mathrm{BR}_2}{\mathrm{BR}_1}\simeq \frac{\Delta_2}{\Delta_1} \gg 1 \,,
\end{align}
the relevant function $x^2 f(x)$ has two maxima of the same height, potentially rendering the mean $x$ useless. This is illustrated in Fig.~\ref{fig:two_decays}.
In such cases, the impact on structure formation should be determined by calculating the transfer function~\cite{Colombi:1995ze,Bezrukov:2014qda,Konig:2016dzg,Murgia:2017lwo}; this is left for future work.

\section{Toy models}
\label{sec:toy_models}

To illustrate the key ideas for the generation of sufficiently cold \unit{keV} DM, let us first discuss a number of toy models. We will mostly restrict ourselves to interactions of scalars and fermions, but will briefly generalize the concepts to interactions with higher-spin particles.

\subsection{Decays involving two particles}
\label{sec:decays_two}

Let us start our discussion with the decay of one real scalar $S$, governed by the Lagrangian (always omitting kinetic terms in the following)
\begin{align}
\L_S = -\tfrac12 m_S^2 S^2-\tfrac12 m_J^2 J^2 - \mu \, S J J\,,
\end{align}
assuming $J$ to be the DM candidate with keV-scale mass $m_J\ll m_S$. If $S$ is in thermal equilibrium with the SM, the dimensionful coupling $\mu$ has to be taken sufficiently small to not thermalize $J$, which will then be produced only through the decay $S\to JJ$~\cite{McDonald:2001vt}. This mechanism yields $\langle p/T\rangle_\mathrm{prod}= 5/2$ (Eq.~\eqref{eq:pT_decay}) and hence a lower mass limit $m_J \gtrsim \unit[12.9]{keV}$ from Eq.~\eqref{eq:translation}, taking the conservative case of production above the electroweak scale. In this simple case one can take into account the quantum BE statistics of the parent particle, obtaining the slightly smaller value $\langle p/T\rangle_\mathrm{prod}= 2.45$ (see~\eqref{eq:pT_decay_BE}) and, correspondingly, the slightly weaker bound $m_J \gtrsim \unit[12.6]{keV}$, which also shows that corrections from quantum statistics are at the few percent level.

The very same discussion can be made for a fermionic DM candidate $N$, replacing $\mu S JJ\to \lambda S \bar{N} N$. 
This has been discussed extensively in the literature~\cite{Shaposhnikov:2006xi,Kusenko:2006rh,Petraki:2007gq,Bezrukov:2014nza,Drewes:2015eoa,Merle:2015oja}.
One obtains, again, just $\langle p/T\rangle_\mathrm{prod}= 5/2$ from our Eq.~\eqref{eq:pT_decay}, if the decays occur when $S$ is still at equilibrium, but larger values if $S$ freezes out before it decays. In all cases above, a \unit[7]{keV} DM candidate is in strong tension with Lyman-$\alpha$ bounds.

A popular alternative to lower the hotness of sterile-neutrino DM produced by particle decays $S\to NN$ is to assume, instead, that $S$ itself is produced via freeze-in~\cite{Merle:2013wta}, i.e.~non-thermally, so that the DM momentum depends on the lifetime of $S$: for a long-lived $S$, the resulting $\langle p/T \rangle$ becomes large, seeing as $p\sim m_S/2$ can be much larger than the temperature when $S$ eventually decays; if the decay is sufficiently fast, on the other hand, values of $\langle p/T \rangle$ down to $1$ become possible, and hence DM masses down to $\approx \unit[5]{keV}$ without violating Lyman-$\alpha$ constraints. While this scenario is certainly attractive, this \emph{double freeze-in} requires both the couplings of $S$ and $N$ to the SM to be very small, $\lesssim 10^{-8}$, thus making the mechanism very hard to probe experimentally. Therefore, we now go beyond this simple scenario.

\subsection{Decays involving three particles}
\label{sec:decays_three}

Extending the particle content gives rise to qualitatively different results. Take three scalars,
\begin{align}
\L_S &= -\tfrac12 m_1^2 S_1^2-\tfrac12 m_2^2 S_2^2-\tfrac12 m_J^2 J^2-\mu_{12} S_1 S_2 J \,,
\end{align}
with mass hierarchy $m_J \ll m_J+m_2 < m_1$. Assuming $\mu_{12}$ to be small enough to not thermalize the DM particle $J$, it will still be produced via the freeze-in decay $S_1\to S_2 J$, with partial width
\begin{align}
\Gamma (S_1\to S_2 J) \ = \ \frac{\mu_{12}^2}{8\pi m_1^2} \, |\vec{p}_J^\mathrm{RF}| \,,
\end{align}
and $J$ momentum in the $S_1$ rest frame (RF)
\begin{align}
|\vec{p}_J^\mathrm{RF}| \simeq \frac{m_1}{2} \left(1-\frac{m_2^2}{m_1^2}\right)\,,
\end{align}
having set $m_J \simeq 0$. For $m_2 \ll m_1$, nothing changes compared to the $S\to JJ$ discussion from above; for $m_2 \to m_1$ on the other hand, the decay rate becomes phase-space suppressed and the momentum of $J$ in the rest frame of $S_1$ goes to zero.
In the thermal-bath frame (TB), the DM momentum is boosted with a $\gamma$ factor $\gamma = E_1/m_1 = (1-v^\mathrm{TB})^{-1/2}$, $v^\mathrm{TB}$ being the velocity of $S_1$. The energy $E_1$ is thermally distributed and thus never much larger than the temperature; in particular, $\gamma$ is not much larger than 1 around the time of DM production. This boost increases the momentum to 
\begin{align}
|\vec{p}_J^\mathrm{TB}| = \gamma |\vec{p}_J^\mathrm{RF}|+ v^\mathrm{TB} \sqrt{m_J^2 + |\vec{p}_J^\mathrm{RF}|^2} \,,
\end{align}
which is parametrically suppressed by $|\vec{p}_J^\mathrm{RF}|$ and $m_J$. The former can be made arbitrarily small for $m_2\to m_1$, while the latter only leaves a tiny contribution $\langle x \rangle \sim m_J/m_1$, fairly irrelevant for the mass scales of interest. This confirms our result from Eq.~\eqref{eq:pT_decay}, meaning the light $J$ actually becomes arbitrarily cold for $m_2 \to m_1$.
In turn, this means Lyman-$\alpha$ constraints can be evaded even for DM particle with mass much below the keV scale, see Eq.~\eqref{eq:translation}.
Taking the concrete case of a \unit[7]{keV} DM particle, we can calculate the necessary mass splitting between $S_1$ and $S_2$ to satisfy the Lyman-$\alpha$ constraints of Eq.~\eqref{eq:7kevLyman} as $m_1>m_2\gtrsim 0.66 m_1$. This can hardly be considered fine-tuning and shows the importance of even a mild phase-space closure.
Staying with this example, the correct relic abundance is obtained for the coupling
\begin{align}
\mu_{12}\simeq \unit[44]{keV}\left(\frac{m_1}{\unit{TeV}}\right)^{3/2} \left\langle \frac{p}{T}\right\rangle_\mathrm{prod}^{-1/2}\,,
\end{align}
assuming $m_1$ sufficiently high above the electroweak scale. An upper limit of order tens of PeV for $m_1$ can be obtained by demanding $\mu_{12}$ to be small enough to not thermalize $J$ via scattering processes.

A similar discussion can be made for decays involving fermions, taking for example the Lagrangian
\begin{align}
\L_F = -\frac{m_S^2}{2} S^2- m_1 \bar{F}_1 F_1 - m_2 \bar{F}_2 F_2 -(\lambda \bar{F}_1 F_2 S+\hc) \,.
\end{align}
For $m_1 \ll m_1+m_2 \lesssim m_S$, the decay $S\to F_1 F_2$~\cite{Kang:2014cia} will produce cold light $F_1$ DM;
for $m_1+m_S \lesssim m_2$, the decay $F_2\to F_1 S$ will produce cold light $F_1$ DM (if $m_1 \ll m_S$) or cold light $S$ DM (if $m_S\ll m_1$).\footnote{Phase-space suppressed decays as a means to obtain colder light DM was also noted in Ref.~\cite{Lello:2014yha} when comparing $\pi \to \mu \,\mathrm{DM}$ and $\pi \to e \,\mathrm{DM}$. The ratio of the resulting free-streaming lengths is precisely the phase-space suppression factor $1- m_\mu^2/m_\pi^2$.}
Since the argument essentially relies purely on kinematics, it does not matter whether the fermions are Majorana or Dirac, nor what the precise Lorentz structure of the coupling is. In fact, this mechanism works for particles of arbitrary spin.

In conclusion, slow decays in which the daughter particles are near the phase-space closure will give a cold spectrum even if one of the daughters has keV mass. Clearly, in a full model there will be competing processes that have to be compared to determine the actual coldness of DM, but the basic idea is incredibly simple and straightforward.

\subsection{Scatterings from scalars}
\label{sec:scatterings_scalars}

As we have seen above, keV particles can be produced arbitrarily cold via phase-space-suppressed \emph{decays}. Extending this idea to phase-space-suppressed \emph{scatterings}, $A B \to C\,\mathrm{DM}$, is not trivial, mainly due to the thermal distribution of the initial particles. We will show nonetheless that it is possible to create cold DM in this way, the degree of coldness being again determined by kinematics.
We restrict ourselves to the interactions of three different particles, as this already features the basic ideas. 

\begin{figure}[t]
\includegraphics[width=0.52\textwidth]{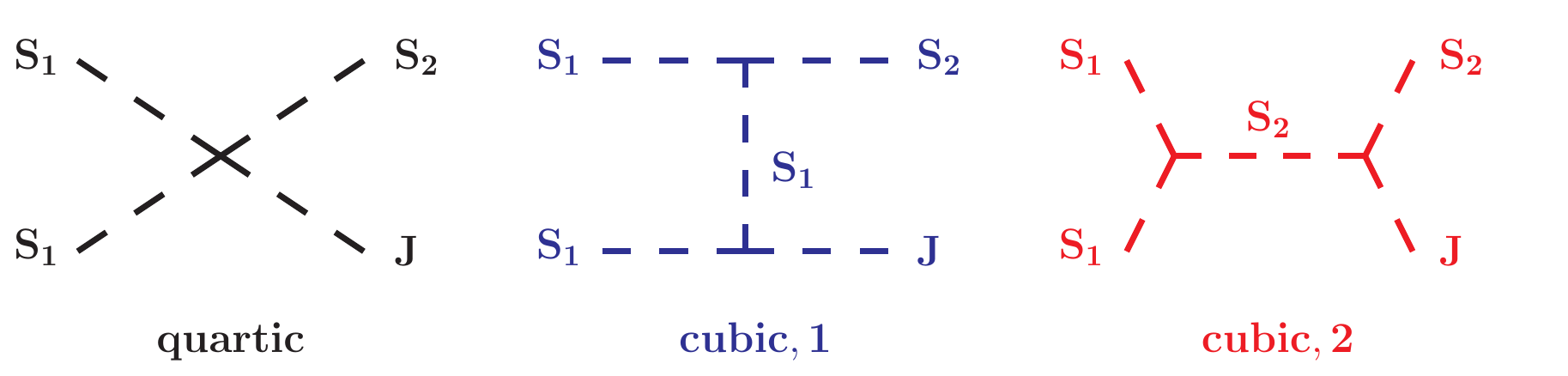}
\caption{
Production processes for $J$ DM via the process $S_1 S_1 \to S_2 J$ for three different models: quartic interaction (Eq.~\eqref{eq:quartic}) and the cubic interactions of Eq.~\eqref{eq:cubic1} and \eqref{eq:cubic2}.
}
\label{fig:three_scalars_diagrams}
\end{figure}

\begin{figure*}[t]
\includegraphics[width=0.49\textwidth]{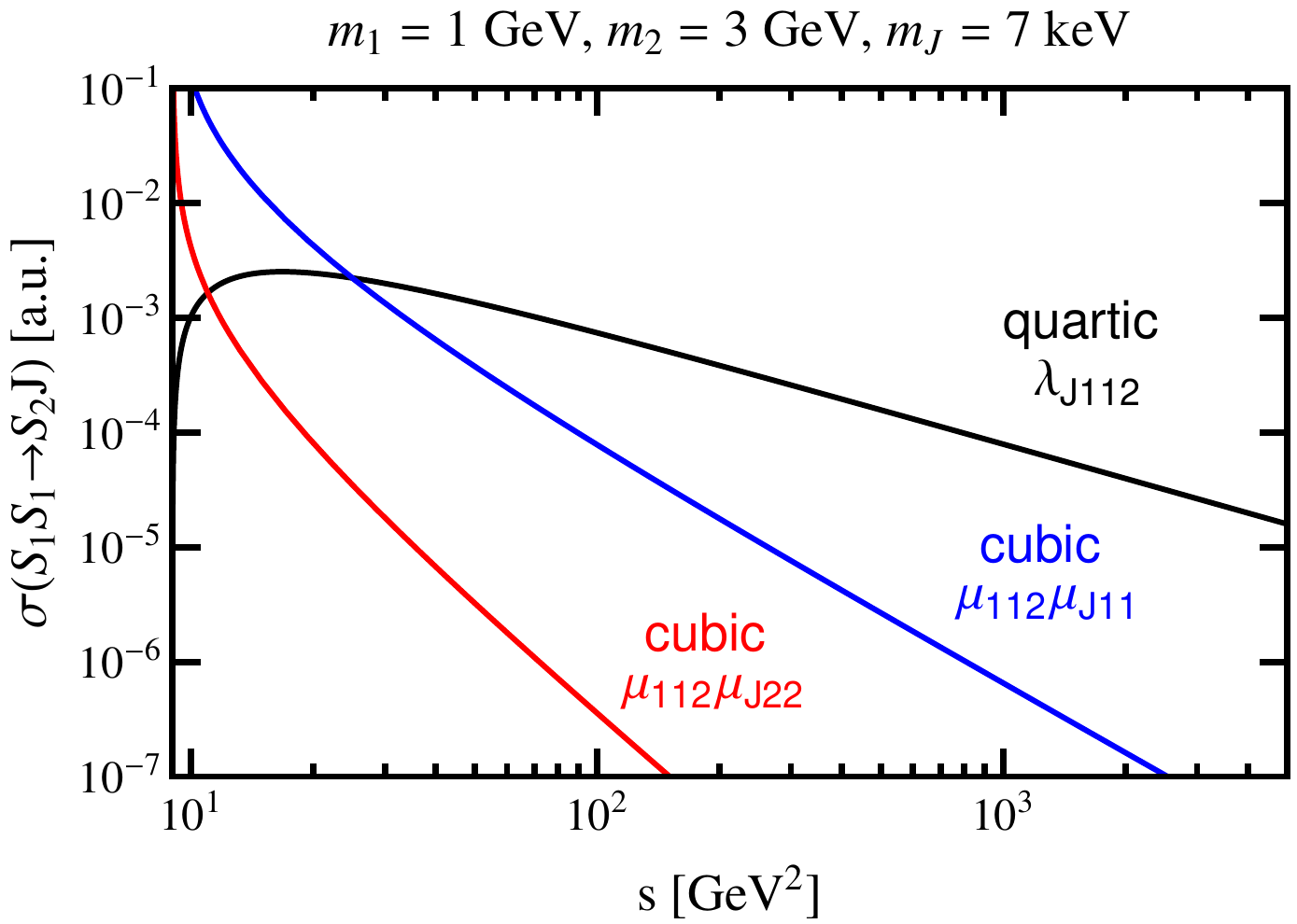} \hspace{1ex}
\includegraphics[width=0.49\textwidth]{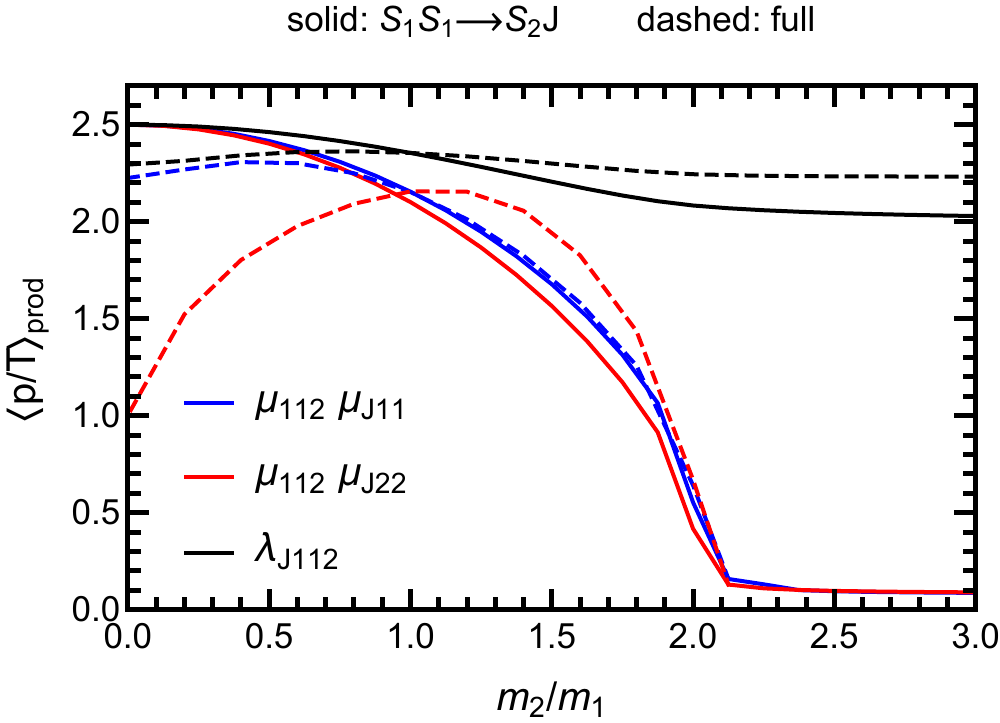}
\caption{
Left: Cross sections $\sigma (S_1 S_1 \to S_2 J)$ obtained from quartic interactions~\eqref{eq:quartic} and two different cubic interactions, Eqs.~\eqref{eq:cubic1}--\eqref{eq:cubic2}.
Right: Resulting $\langle p/T\rangle_\mathrm{prod}$ for these three processes as a function of the $S_{1,2}$ mass ratio $m_2/m_1$. The dashed lines are the results including also the $S_1 S_2 \to S_1 J$ process, active when $S_2$ is also thermalized.\label{fig:three_scalars_xsecs_2}
}
\end{figure*}

Let us focus on the pure scalar case, i.e.~three real scalars $S_{1}, S_{2}, J$, with one quartic interaction term,
\begin{align}
\L_{\mathrm{quartic}} = -\frac{m_1^2}{2} S_1^2  -\frac{m_2^2}{2} S_2^2  -\frac{m_J^2}{2} J^2  - \lambda_{J112} J S_1^2 S_2 ,
\label{eq:quartic}
\end{align}
and DM candidate $J$, $m_J\ll m_2\leq 2m_1$. Assuming that $S_1$ and $S_2$ are in thermal equilibrium with the SM, $J$ is produced by the freeze-in scatterings $S_1 S_1\to S_2 J$ (Fig.~\ref{fig:three_scalars_diagrams}) and $S_1 S_2\to S_1 J$. The cross sections for these processes are given in App.~\ref{sec:xsecs} and can be readily used to calculate the DM distribution function $f$ via Eq.~\eqref{eq:eq_scat} and the average momentum via Eq.~\eqref{eq:mean_pT}. For a keV-range $m_J$, one finds a warm spectrum, $\langle p/T\rangle \sim 2$, independently of the $S_{1,2}$ masses (Fig.~\ref{fig:three_scalars_xsecs_2}). 
This conclusion also holds for processes such as $S_i S_j \to J J$: quartic interactions can not give cold keV DM and have to be suppressed, with couplings at most $\mathcal{O}(10^{-8})$ for $m_{S_{i,j}} \sim \unit[100]{GeV}$.

With \emph{quartic} interactions sufficiently suppressed, let us focus on two sets of some \emph{cubic} interactions,
\begin{align}
\begin{split}
\L_{\mathrm{cubic},1} &=  -\tfrac{1}{2} m_1^2 S_1^2  - \tfrac{1}{2} m_2^2 S_2^2  -\tfrac{1}{2} m_J^2 J^2  \\ 
&\quad- \mu_{112} \,S_1^2 S_2- \mu_{J11} \,J S_1^2  \,,
\end{split}
\label{eq:cubic1}
\end{align}
and
\begin{align}
\begin{split}
\L_{\mathrm{cubic},2} &= -\tfrac{1}{2} m_1^2 S_1^2  -\tfrac{1}{2} m_2^2 S_2^2  -\tfrac{1}{2} m_J^2J^2  \\
&\quad  - \mu_{112} \, S_1^2 S_2 - \mu_{J22} \,J S_2^2 \,.
\end{split}
\label{eq:cubic2}
\end{align}
The full DM-production cross sections for all cases are given in Appendix~\ref{sec:xsecs}, assuming for simplicity that $\mu_{Jii}\ll \mu_{112}$ in order to neglect $S_i S_j\to JJ$ processes which lead to rather warm DM.
The channel of interest to produce the light DM candidate $J$ is $S_1 S_1\to S_2 J$, either via $t$ or $s$-channel (see Fig.~\ref{fig:three_scalars_diagrams}). In Fig.~\ref{fig:three_scalars_xsecs_2} we compare the $S_1 S_1 \to S_2 J$ cross sections with quartic and cubic interactions; the cubic interactions decrease much faster than the quartic one for large $s$, 
\begin{align}
\sigma_{\mathrm{quartic}} \to \frac{1}{s}\,, &&
\sigma_{\mathrm{cubic},1} \to \frac{1}{s^2}\,, &&
\sigma_{\mathrm{cubic},2} \to \frac{1}{s^3}\,,
\end{align}
and are dominated by low $s$ close to the phase-space closure, $s\simeq \text{max} \{m_2^2,4 m_1^2\}$. Writing $s= (p_2 + p_J)^2 = m_2^2 + \mathcal{O}(p_J)$, it is then clear that the cubic cross sections are dominated by small $J$ momentum $p_J \ll p_2$ if $m_2^2 > 4 m_1^2$, which just means that the emission of a cold $J$ is preferred. This is shown in the right panel of Fig.~\ref{fig:three_scalars_xsecs_2}, where the solid lines denote $\langle p/T \rangle_{\rm prod}$ for the quartic and cubic cases above, as function of the mass ratio $m_2/m_1$.

As can be seen in Fig.~\ref{fig:three_scalars_xsecs_2} (right), all $S_1 S_1\to S_2 J$ scattering processes lead to $\langle p/T \rangle_{\rm prod} = 5/2$ in the limit $m_2\ll m_1$, but a non-zero $m_2$ cools down the DM due to phase space effects. 
For the cubic interactions, $m_2> 2 m_1$ even leads to $\langle p/T \rangle_{\rm prod} \simeq 0.1$, i.e.~very cold light DM!
This allows for DM masses down to $\unit[0.5]{keV}$ without violating Lyman-$\alpha$ constraints (Eq.~\eqref{eq:translation}).

The reason for this surprising behaviour can be traced back to the low-$s$ peak of the corresponding cross sections (Fig.~\ref{fig:three_scalars_xsecs_2} (left)), but can also be understood at the amplitude level as \emph{near resonances}.
Looking, for example, at the $t$-channel amplitude for $S_1 S_1\to S_2 J$ via $S_1^2 S_2$ and $S_1^2 J$ couplings (Fig.~\ref{fig:three_scalars_diagrams} (middle)),
\begin{align}
\mathcal{M}\propto \frac{\mu_{112} \,\mu_{J11}}{m_1^2 - (p_1 - p_J)^2} \,,
\label{eq:resonance}
\end{align}
it is clear that the exchanged particle $S_1$ goes on-shell for $p_J\to 0$, signalling an infrared divergence. Energy--momentum conservation at the second vertex requires $m_2 > 2 m_1$ to allow $S_1$ to be on shell, and one can identify the divergence as soft for $m_J\to 0$ rather than collinear. For a non-zero $m_J$, the virtual particle can never actually hit the resonance, but comes closer and closer the smaller $\vec{p}_J$ becomes. Thus, the production of \emph{cold} light $J$ is preferred quite dramatically as a result of a near $t$-channel resonance.
For the other cubic coupling (Fig.~\ref{fig:three_scalars_diagrams} (right)), the process is enhanced by a near $s$-channel resonance, with the virtual $S_2$ getting closer to the resonance the smaller $\vec{p}_J$ becomes.

An alternative interpretation of Fig.~\ref{fig:three_scalars_xsecs_2} (right) is that a vanishing DM momentum $p_J = 0$ reduces the cubic scattering $S_1 S_1 \to S_2 J$ to an inverse decay $S_1 S_1\to S_2$, provided that $m_2 > 2 m_1$. This $2\to 1$ process is parametrically fast compared to the $2\leftrightarrow 2$ scattering one has for the emission of a \emph{hard} $J$. For $m_2 > 2 m_1$, phase space thus prefers to emit cold light DM, whereas $m_2 < 2 m_1$ always leads to $2\to 2$ scatterings with rather warm DM.
%Extending the particle content by another scalar, one can then immediately write down processes of interest by attaching $J$ to any of the external particles in the inverse decay $S_1 S_2 \to S_3$ via cubic $S_j^2 J$ coupling, and choosing the mass hierarchy $m_J\ll m_1 + m_2<m_3$. Since this is qualitatively similar to the situation with just two massive scalars, we will not discuss this further.

We have identified $S_1 S_1 \to S_2 J$ as a promising process to produce cold light DM with cubic interactions; to not thermalize $J$, we expect a coupling hierarchy $\mu_{112}\gg \mu_{Jii}$, which then typically puts $S_2$ in equilibrium. This opens up the scattering channel $S_1 S_2\to S_1 J$, which could easily endanger the coldness of $J$. The results including this process, assuming thermalized $S_1$ and $S_2$, are shown by dashed lines in Fig.~\ref{fig:three_scalars_xsecs_2}. We see that the new process plays no role for the cubic interactions in the region of interest $m_2 > 2 m_1$, still giving cold DM. For smaller $m_2$, the new process actually helps to cool down the momentum. This effect is strongest for the cubic interactions of Eq.~\eqref{eq:cubic2} (red in Fig.~\ref{fig:three_scalars_xsecs_2}), as the $t$-channel exchange exhibits both a near soft and a near collinear divergence for $m_J,m_2\to 0$, leading to $\langle p/T\rangle_{\rm prod} =1$. No collinear divergence is obtained for the second (Compton-like) cubic process (blue in Fig.~\ref{fig:three_scalars_xsecs_2}), which is then more inefficient in cooling down the DM.

As far as the time of DM production goes, we can interpret the center-of-mass energy $s$ in Fig.~\ref{fig:three_scalars_xsecs_2} (left) as the bath temperature, which decreases with time. For the cubic interactions with $m_2 > 2 m_1$, the highly peaked cross sections then ensure that DM will be produced almost resonantly around the temperature $T\sim m_2/3$. The relic abundance in the cold regime of interest, $2 m_1\ll m_2$, is then approximately given by
\begin{align}
\Omega_\mathrm{DM}h^2 &\simeq 3.4 \left(\frac{m_\mathrm{DM}}{\unit[7]{keV}}\right) \left(\frac{\unit{TeV}}{m_2}\right)^3 \left(\frac{\unit{GeV}}{m_1}\right)^2 \left(\frac{\mu_{112}\mu_{J11}}{(\unit[10]{MeV})^2}\right)^2 \!\! ,\\
\Omega_\mathrm{DM}h^2 &\simeq 1.5 \times 10^{-2} \left(\frac{m_\mathrm{DM}}{\unit[7]{keV}}\right) \left(\frac{\unit{TeV}}{m_2}\right)^5 \left(\frac{\mu_{112}\mu_{J22}}{(\unit[100]{MeV})^2}\right)^2 \!\! ,
\end{align}
for the two different cubic interactions. Here, we assumed a DM production time above the electroweak scale and also the hierarchy $\mu_{Jii}\ll \mu_{112}$.

All in all, we can see from Fig.~\ref{fig:three_scalars_xsecs_2} (right) that light DM can be produced with $\langle p/T\rangle_{\rm prod} =2.5$--$\mathcal{O}(0.1)$ depending on the mass spectrum of the additional scalars. For very cold DM, the requirement for the cubic couplings is that $p_\mathrm{DM}\to 0$ leads back to a valid inverse decay process.

\subsection{Scatterings from fermions}
\label{sec:scatterings_fermions}

Fig.~\ref{fig:three_scalars_xsecs_2} (right) illustrates how to obtain cold light DM in the scatterings of \emph{scalars} by picking the right coupling structure and mass hierarchy. Contrary to the decay case discussed in Sec.~\ref{sec:decays_three}, the extension of this mechanism to particles with non-zero spins is not trivial.
Consider for example the interactions of a Dirac fermion $\Psi$ with a scalar $S$ and a pseudoscalar $A$:
\begin{align}
\begin{split}
\L_\Psi &= -\tfrac12 m_S^2 S^2 -\tfrac12 m_A^2 A^2 - m_\Psi \overline{\Psi}\Psi \\
&\quad- \overline{\Psi}(y_S S + y_A A \,i \gamma_5)\Psi \,.
\end{split}
\end{align}
For $m_S\ll  2 m_\Psi < m_A$, one obtains a very cold $\langle p/T\rangle = \mathcal{O}(0.1)$ for $S$ from the inverse-decay-like process $\overline{\Psi}\Psi\to S A$, in complete analogy to the pure scalar case discussed above. Flipping $S\leftrightarrow A$ changes the picture dramatically, though: for $m_A \ll 2 m_\Psi < m_S$, the process $\overline{\Psi}\Psi\to S A$ does \emph{not} produce cold $A$, but rather $\langle p/T\rangle \simeq 2$. This can be understood at the amplitude level, which for \emph{soft} $S=\mathrm{DM}$ takes the form
\begin{align}
\begin{split}
\mathcal{M} &\simeq y_S y_A \, \overline{v}(p_{\overline{\Psi}}) i \gamma_5 \frac{\slashed{p}_\Psi-\slashed{p}_\mathrm{DM} + m_\Psi }{m_\Psi^2 - (p_\Psi- p_\mathrm{DM})^2} u (p_\Psi)\\
&\overset{p_\mathrm{DM}\to 0}{\to} y_S y_A \, \overline{v}(p_{\overline{\Psi}}) i \gamma_5 \frac{2 m_\Psi }{2 p_\Psi  p_\mathrm{DM} -  p_\mathrm{DM}^2} u (p_\Psi) \,,
\end{split}
\end{align}
whereas a \emph{soft} $A=\mathrm{DM}$ is heavily suppressed by $p_\mathrm{DM}/m_\Psi$ in comparison,
\begin{align}
\begin{split}
\mathcal{M} &\simeq y_S y_A \, \overline{v}(p_{\overline{\Psi}})\frac{\slashed{p}_\Psi-\slashed{p}_\mathrm{DM} + m_\Psi }{m_\Psi^2 - (p_\Psi- p_\mathrm{DM})^2}  i \gamma_5  u (p_\Psi)\\
&\overset{p_\mathrm{DM}\to 0}{\to} y_S y_A \, \overline{v}(p_{\overline{\Psi}}) i \gamma_5 \frac{\slashed{p}_\mathrm{DM}}{2 p_\Psi  p_\mathrm{DM} -  p_\mathrm{DM}^2} u (p_\Psi) \,.
\end{split}
\end{align}
The amplitude above does not vanish in the soft limit due to the competing propagator divergence, resulting in a finite expression~\cite{Weinberg:1996kr}. Our cubic interactions thus evade Adler's zero~\cite{Adler:1964um}, which would otherwise lead to a vanishing amplitude for the emission of one soft Goldstone boson. It is not relevant for us whether the amplitude goes to zero or not, the main point here is that the emission of a soft pseudoscalar is suppressed compared to a soft scalar.

On a conceptual level, one can understand this by noting that $\overline{\Psi}\Psi\to S A$ does not reduce to the inverse decay $\overline{\Psi}\Psi\to S$ no matter how soft $A$ is, because the $\gamma_5$ changes the spin of the involved $\Psi$, so $\overline{\Psi}\Psi\to S A$ is always a true $2\to 2$ process that does not enjoy the $2\to 1$ enhancement we found for a light cold $S$.
This argument goes through for other processes as well; for example, emitting a light scalar from an inverse decay via the CP-even coupling to vector bosons $S F_{\mu\nu} F^{\mu\nu}$ gives indeed a cold $S$, whereas the emission of a light pseudoscalar $A$ via  $A F_{\mu\nu}\tilde F^{\mu\nu}$ does not give a cold~$A$. 

The above discussion has the far reaching implication that \emph{it is not possible to create a cold light pure pseudoscalar via scatterings}.\footnote{A loophole to the statement above can be found in the scalar case by extending the particle content: if the DM interaction is $J S_1 S_2$ with \emph{degenerate} $S_{1,2}$, there is no decay and one can once again create cold DM by attaching $J$ to an inverse decay, e.g.~$S_3 S_3\to S_1$, so that $S_3 S_3\to S_2 J$ with spectrum $m_J \ll 2 m_3 < m_1 = m_2$. If $S_{1,2}$ are merely the real and imaginary part of one complex scalar, the interaction $J S_1 S_2$ does not violate CP for a pseudoscalar $J$ and we have managed to create a cold pure pseudoscalar via scattering. Note that the $J S_1 S_2$ coupling breaks the $U(1)$ symmetry $J$ is a Goldstone of, otherwise $S_1$ and $S_2$ could not be degenerate and the main production channel would again be decay, e.g.~$S_2\to S_1 J$. \smallskip}
This is of course unfortunate considering that a major motivation for light bosonic DM comes in the form of pseudo-Goldstone bosons, i.e.~pseudoscalars, whose mass is protected by a global symmetry. Nevertheless, depending on the UV completion of the model, one can of course have CP violation that endows the pseudoscalar with some CP-even coupling, which can then be used to produce cold DM. This is precisely the route that we will follow in Secs.~\ref{sec:heavy_scalar} and~\ref{sec:light_scalar}.

\subsection{Scatterings with vector DM}
\label{sec:scatterings_vectorDM}

At last, let us mention spin-1 DM. From the arguments above, it is clear that a light vector boson $V$ can be made cold by emitting it from an inverse decay, as long as the diagonal coupling of the soft $V$ does not change the spin of the heavy particle. This is the case, for example, for a $\overline{\Psi}\gamma_\mu \Psi$ coupling to Dirac fermions,\footnote{Coupling $V$ to an axial-current leads to a spin-flip amplitude for soft $V$ and hence no cold $V$, analogous to the pseudoscalar case discussed above.} as can be verified using the Gordon decomposition identity in the limit of small momentum transfer,
\begin{align}
\overline{u}(p) \gamma^\mu u(p+q) =\frac{p^\mu}{m_\Psi}\overline{u}(p) u(p) + \mathcal{O}\left(\frac{q}{m_\Psi}\right)\,.
\end{align}
 In this case, $V$ is produced essentially via initial- or final-state radiation, both of which are famously infrared divergent. It is precisely this divergence that makes this process dominant and leads to a cold $V$. In contrast to QED, our DM vector boson is necessarily massive, so the divergence is automatically regulated; furthermore, the DM coupling is required to be tiny, in particular compared to Sudakov logs such as $\log (m_\mathrm{heavy}/m_\mathrm{DM})$, so there is no need to calculate loop corrections or sum over additional soft DM emissions.

\subsection{What we have learned}

Let us briefly summarize the main lessons so far to obtain cold light DM from decays and scatterings in a more general setup:
\begin{itemize}
	\item The decay $A\to B\,\mathrm{DM}$ of a thermalized $A$ can lead to DM momenta from $\langle p/T\rangle_{\rm prod} =2.5$ down to tiny $\mathcal{O}(m_\mathrm{DM}/m_A)$, depending only on $m_B/m_A$. The spins of the particles involved do not matter, nor does the Lorentz structure of the coupling.
	\item The scattering process $A\, B \to C\, \mathrm{DM}$ can lead to $\langle p/T\rangle_{\rm prod} =2.5$ down to $\mathcal{O}(0.1)$. At the lower end, this requires a mass hierarchy $m_A+m_B < m_C$ and \emph{bosonic} DM with a cubic coupling to either $A$, $B$ or $C$ that is diagonal in spin space (at least for small momentum transfer). In the presence of more particles, this cubic coupling could also be to two different, but degenerate, particles.
	\end{itemize}
An alternative formulation can be given as follows: one can obtain cold DM from scattering processes that have a soft infrared divergence in the limit $p_\mathrm{DM}, m_\mathrm{DM}\to 0$.
Despite this near divergence, the DM production is actually not sensitive to the light DM mass due to the thermal environment.

\section{Application to full models}
\label{sec:full_models}

Having identified promising coupling structures to produce cold keV-scale DM via decays and scatterings, let us connect them to some popular models. We focus here to the case of (pseudo) scalar \unit{keV}-scale DM $J$. Differently from the case of sterile right-handed neutrinos, in the pseudoscalar case the lightness of keV DM can originate dynamically, rather than being imposed \emph{by hand}, from $J$ being the pseudo-Goldstone boson of an approximate continuous symmetry~\cite{Frigerio:2011in}. Examples for this are the axion (Peccei--Quinn $U(1)$)~\cite{Peccei:1977hh,Weinberg:1977ma,Wilczek:1977pj}, majoron (lepton number $U(1)_L$ or $U(1)_{B-L}$)~\cite{Chikashige:1980ui,Schechter:1981cv}, and familons (lepton family number symmetries)~\cite{Wilczek:1982rv,Grinstein:1985rt}.
Recently, relaxion models have been put forward that could also be taken as a motivation for a new pseudoscalar~(see for example Refs.~\cite{Graham:2015cka, Flacke:2016szy}).
Motivated by the tantalizing hint for a \unit[3.5]{keV} line we fix the mass of $J$ to \unit[7]{keV} in this section. In Sec.~\ref{sec:decay} we will then discuss how the line could be generated by the radiative decay of the pseudo-Goldstone boson $J$.

\subsection{Right-handed neutrino decay in the singlet majoron model: \texorpdfstring{$N_1\to N_2 J$}{N1 to N2 J}}
\label{sec:RHNdecay}

To produce cold keV-scale DM from a decay $A\to B\, \mathrm{DM}$ requires an off-diagonal coupling of DM to two moderately degenerate  heavy particles $A$ and $B$. Since the SM does not provide any useful candidates for $A$ and $B$, new heavy particles have to be introduced. A simple example comes in the form of singlet-majoron models~\cite{Chikashige:1980ui,Schechter:1981cv}, where heavy right-handed neutrinos $N$ are introduced to generate seesaw-suppressed neutrino masses and the majoron mass itself is protected by $U(1)_{B-L}$ and thus naturally small. The majoron's only tree-level couplings are to neutrinos, so the $N_i \to N_j J$ decays of the heavy neutrinos seem promising for our purpose. 
Explicitly, the partial widths of neutrino decays into a (massless) majoron are
\begin{align}
\Gamma (N_i \to \nu_j J) &\simeq \frac{M_{N,i}^3}{16\pi f^2} \left| \Theta^\dagger_{ij}\right|^2 ,\label{eq:NnuJ_decay}\\
\Gamma (N_i \to N_j J) &\simeq \frac{M_{N,i}^3}{16\pi f^2} \left| (\Theta^\dagger \Theta)_{ij}\right|^2  \left(1-\frac{M_{N,j}^2}{M_{N,i}^2}\right)^3\hspace{-1ex},\label{eq:NNJ_decay}\\
\Gamma (\nu_i \to \nu_j J) &\simeq \frac{M_{\nu,i}^3}{16\pi f^2} \left| (\Theta \Theta^\dagger)_{ij}\right|^2 \left(1-\frac{M_{\nu,j}^2}{M_{\nu,i}^2}\right)^3 \hspace{-1ex},\label{eq:nunuJ_decay}
\end{align}
with $\Theta \equiv U^\dagger m_D M_N^{-1}$ and the $B-L$ breaking scale $f$. 
The decay of active neutrinos in Eq.~\eqref{eq:nunuJ_decay} has already been discussed long ago in Ref.~\cite{Schechter:1981cv}, where it was realized that the rates are unobservably small, but this is of course not relevant for the keV majoron that we have in mind. 
Of interest for us are only the $N_i \to N_j J$ decays of Eq.~\eqref{eq:NNJ_decay}, which are at first sight less suppressed; however, since the matrix $\Theta$ corresponds to the small seesaw expansion parameter, it is clear that $\Gamma (N_i \to N_j J) \ll \Gamma (N_i \to \nu J)$, so that only a tiny fraction of the keV majoron DM can be produced \emph{cold}.
To be more precise, electroweak precision data conservatively restricts the entries of $\Theta$ to be below $0.1$~\cite{Antusch:2015mia}, which is enough to suppress the cold DM part below the percent level.
We are then precisely in the scenario of Fig.~\ref{fig:two_decays}, where hot and cold DM are mixed and require a more sophisticated discussion than just looking at $\langle p/T\rangle$. An evaluation of the Lyman-$\alpha$ bounds on this mixed scenario has to be left for future work.

Let us nevertheless comment on DM production via $N\to\nu J$ decay, with $N$ in equilibrium, which has $\langle p/T\rangle = 2.5$ and thus works for $m_J\gtrsim \unit[13]{keV}$. Summing over all neutrino channels and replacing $f$ by the unavoidable DM decay rate $\Gamma (J\to\nu\nu)$~\cite{Garcia-Cely:2017oco} (see Eq.~\eqref{eq:majoron_to_neutrinos}), we obtain an expression that is independent of $m_J$:
\begin{align}
\Omega h^2 \simeq 0.13 \left(\frac{\unit[10^{18}]{s}}{\tau_{J\to\nu\nu}}\right) \frac{(\unit[0.1]{eV})^2}{\sum m_\nu^2}\,\frac{\tr (m_D M_N^{-1} m_D^\dagger )}{\unit{MeV}}\,.
\end{align}
A sufficiently stable DM abundance from $N\to\nu J$ thus requires $\tr (m_D M_N^{-1} m_D^\dagger )\sim \unit{MeV}$, while keeping
\begin{align}
|\tr (m_D M_N^{-1} m_D^T )| \leq \sum_\nu m_\nu < \unit{eV}\,.
\end{align}
This is possible, but requires fine-tuned matrix cancellations and typically Yukawa couplings close to the perturbativity limit. For $m_J > \unit{MeV}$, the lifetime needs to be even longer in order to satisfy limits from searches for monochromatic neutrinos~\cite{Garcia-Cely:2017oco}, thus requiring an even larger trace.
We will not discuss this case further.

\subsection{Right-handed neutrino decay in extended majoron models}
\label{sec:RHNdecay_extended}

As we have seen above, the standard singlet-majoron model is probably incapable of producing sufficiently cold keV-scale majoron DM. It is however straightforward to identify minor modifications that are viable, i.e.~models where $J$ has a larger off-diagonal coupling to two heavy right-handed neutrinos. A sufficient condition for this is $[M_N, \lambda]\neq 0$, where $M_N$ is the mass matrix of $N$ and $\lambda$ the coupling to the majoron. In the singlet majoron model, one could achieve this by simply introducing some \emph{explicit} $U(1)_L$ breaking terms in $M_N$ together with the spontaneously generated entries. Since lepton number is then no longer a good symmetry of the Lagrangian, the majoron also acquires a mass that can be expressed in terms of the neutrino masses~\cite{Frigerio:2011in}. Depending on the structure of the explicit $U(1)_L$ breaking terms, it is in some cases still possible to identify a \emph{family symmetry} $U(1)_X$, which upon spontaneous breaking gives rise to a massless \emph{familon}, which then plays the role of DM for us. Switching $U(1)$ lepton number for some family number to obtain larger off-diagonal couplings is actually a rather old idea that was developed in the context of neutrino decay~\cite{Gelmini:1983ea}. Let us outline a couple of interesting cases:

\subsubsection{Flavour symmetries}

As is well known, the SM has the global symmetry group
	\begin{align}
	\begin{split}
\qquad	U(1)_{B+L}&\times U(1)_{B-L} \\
	&\times U(1)_{L_\mu-L_\tau}\times U(1)_{L_\mu+ L_\tau -2 L_e}\,,
	\end{split}
	\end{align}
	of which $B+L$ is broken at the non-perturbative level~\cite{'tHooft:1976up} but the remaining $U(1)^3$ is anomaly free once three right-handed neutrinos are introduced~\cite{Araki:2012ip}. For the standard majoron model, one breaks $B-L$ spontaneously and the other two $U(1)$ symmetries explicitly. Since this is not useful to obtain cold keV pseudo-Goldstone DM, let us instead break a different linear combination $U(1)_X \subset U(1)_{B-L} \times U(1)_{L_\mu-L_\tau}\times U(1)_{L_\mu+ L_\tau -2 L_e}$ spontaneously and the orthogonal two explicitly. Even if only one complex scalar is introduced to break the $U(1)_X$, there are many choices that lead to a viable neutrino mass matrix; some cases (e.g.~$X= B+L_e-3 L_\mu - L_\tau$) even feature testable texture zeros~\cite{Araki:2012ip}.
	The three choices $X = L_\mu-L_\tau$, $B- 3 L_e$, and $B+ 3 (L_e- L_\mu-L_\tau)$ are of particular interest, because they correspond to the symmetries in $M_\nu$ for quasi-degenerate, normal, and inverted neutrino hierarchies, respectively~\cite{Choubey:2004hn,Heeck:2012cd}. For example, the Dirac mass matrix $m_D$ is diagonal in the $L_\mu-L_\tau$ case, while the right-handed neutrinos have a mass matrix that consists of one bare mass term $M_N^\mathrm{sym}$ and one term that arises from the vacuum expectation value of a $L_\mu - L_\tau = 1$ scalar $\sigma_1 = \langle \sigma_1\rangle +(\sigma^0_1 + i J)/\sqrt{2}$,
	\begin{align}
	\begin{split}
	\qquad M_N &= M_N^\mathrm{sym} + \langle \sigma_1\rangle \lambda\\
	&= \matrixx{A & 0 & 0\\ 0 & 0 & B\\ 0 & B & 0} + \langle \sigma_1\rangle \matrixx{0 & \lambda_1 & \lambda_2\\\lambda_1 & 0 & 0\\ \lambda_2 & 0 & 0} ,
	\end{split}
	\end{align}
	resulting in an active-neutrino mass matrix $M_\nu \simeq -m_D M_N^{-1} m_D^T$ with two vanishing minors~\cite{Araki:2012ip} and typically quasi-degenerate neutrinos~\cite{Crivellin:2015lwa}. The off-diagonal couplings of the $L_\mu-L_\tau$ Goldstone boson $J$ to the heavy neutrinos are then given by $\lambda_j$, whereas the decay $N\to\nu J$ depends on $m_D$. The latter can thus be suppressed, making $N_i \to N_j J$ the dominant (cold) DM production channel. Note that two of the right-handed neutrinos are degenerate in the limit $M_N^\mathrm{sym} \gg \langle \sigma_1\rangle$, i.e.~form a pseudo-Dirac pair, which makes them ideal candidates for the phase-space suppressed $N_1 \to N_2 J$ decays we want to exploit for cold DM.
	
	As a second example, let us look at the global symmetry $B+ 3 (L_e- L_\mu-L_\tau)$, where one has again a bare mass term that is $U(1)'$ invariant and a term that is induced by a scalar $\sigma_6 = \langle \sigma_6\rangle +(\sigma^0_6 +i J)/\sqrt{2}$ with $U(1)'$ charge~$6$,
		\begin{align}
	\begin{split}
\qquad	M_N &= M_N^\mathrm{sym} + \langle \sigma_6\rangle \lambda\\
	&= \matrixx{0 & A & B\\ A & 0 & 0\\ B & 0 & 0} + \langle \sigma_6\rangle \matrixx{\lambda_1 & 0 & 0\\ 0 & \lambda_2 & \lambda_3\\ 0 & \lambda_3 & \lambda_4} .
	\end{split}
	\end{align}
	This leads to an approximately $L_e$ symmetric $M_\nu \simeq - m_D M_N^{-1} m_D^T$ in the limit $M_N^\mathrm{sym} \gg \langle \sigma_6\rangle$ and thus normal hierarchy~\cite{Heeck:2012cd}. The discussion is completely analogous to the $L_\mu-L_\tau$ case above: the Goldstone boson $J$ can once again have large off-diagonal couplings $\lambda$ to the heavy mass eigenstate neutrinos. Here, two of the right-handed neutrinos are again naturally degenerate, while the third one is considerably lighter, being massless in the $U(1)'$ symmetric limit. All the ingredients for cold keV pseudo-Goldstone DM are present, nicely incorporated into a viable and well-motivated flavour-symmetric model.
	
\subsubsection{Extended seesaw mechanisms}
Instead of charging SM families under the global $U(1)$ that leads to the Goldstone boson of interest, one can also introduce more singlet fermions and identify a family charge among them. A well known example here is the inverse seesaw mechanism, used early on to enhance neutrino decay~\cite{GonzalezGarcia:1988rw}. To avoid confusion, it is convenient to denote some of the right-handed singlet fermions by $N$ and some by $S$. The inverse seesaw mass matrix in the basis $(\nu_L^c, N,S)$ is then of the form
	\begin{align}
\qquad	\mathcal{M}^\mathrm{ISS} = \matrixx{0 & m_D & 0\\ m_D^T & 0 & M \\ 0 & M^T & \mu} ,
	\end{align}
	with $\mu \ll m_D\ll M$. Similar to the flavor-symmetric cases discussed above, the inverse seesaw provides quasi-degenerate heavy mass eigenstates, perfectly suited for our purpose. In the model of Ref.~\cite{GonzalezGarcia:1988rw} it is the $\mu$ entry that is generated by spontaneous $U(1)$ breaking and hence provides the majoron coupling; in simpler models, one would also generate a small $\mathcal{M}_{22}^\mathrm{ISS}$ entry, without much impact on the resulting phenomenology. Explicitly, one can assign a $B-L$ charge $+1$ to $S$ to obtain the leading order $\mathcal{M}^\mathrm{ISS}$, with small $\mathcal{M}_{22}^\mathrm{ISS}$ and $\mathcal{M}_{33}^\mathrm{ISS}$ being induced spontaneously from a scalar VEV with $B-L = 2$.
	
This is not the only charge assignment of interest for us; picking instead $(B-L)(S) = 0$ and a scalar $\sigma_1$ with $(B-L)(\sigma_1)=1$ gives $\mathcal{M}^\mathrm{ISS}$ from above, only that $\mu$ is now a bare mass term and $M\propto \langle \sigma_1 \rangle$, so the majoron couplings are given by a different term. 
These are of course only the simplest cases of interest, with many more conceivable scenarios.

\hspace{3ex}

The above discussion should be sufficient to illustrate the necessary model building to obtain cold keV-scale pseudo-Goldstone DM from the decays of right-handed neutrinos. In flavour symmetric scenarios or extended seesaw model one can easily make $N_i \to N_j J$ the dominant decay mode, thus ensuring that the resulting keV-scale DM will be sufficiently cold and abundant.
A detailed quantitative study of these scenarios will be given elsewhere, seeing as it requires a dedicated analysis of active neutrinos and lepton flavour violation as well.

\subsection{Heavy mediator scalar: \texorpdfstring{$H\bar H\to SJ$}{HH to SJ}}
\label{sec:heavy_scalar}

We now discuss how to obtain a cold-enough \unit[7]{keV} DM $J$ by making use of the scattering processes discussed in Sec.~\ref{sec:scatterings_scalars}. For production above the EWPT, the minimal way to do this is to make use of the Higgs boson $H$, a real singlet $S$ and the DM $J$, the latter potentially being the pseudo-Goldstone boson of a $U(1)$ symmetry. Notice that this model has the same number of beyond-the-SM degrees of freedom as the model with right-handed neutrino DM $N$ produced by $S \to N N$ decays, discussed in Section~\ref{sec:decays_two}. Remarkably, the main advantage here is that, differently from the double freeze-in $S \to NN$ model, where all the BSM fields are feebly coupled to the SM, here $S$ is required to have large couplings with the Higgs doublet, so that the mechanism yields interesting phenomenology at current and future experiments. Moreover, as mentioned above, the lightness of a pseudoscalar $J$ can originate dynamically from a Nambu--Goldstone mechanism.

Therefore, let us mimic Eq.~\eqref{eq:cubic2} by setting $S_1 \to H$, $S_2 \to S$, $J$ still being our DM candidate,
\begin{align}\label{eq:lagr_heavy}
\mathcal{L} &=  - \frac{m_S^2}{2} S^2 - \frac{ m_J^2}{2} J^2 - \mu_{HHS}  |H|^2 S - \mu_{SSJ} S^2 J\,.
\end{align}
If $J$ is the Goldstone boson of some $U(1)$ symmetry, the parameters $m_J$ and $\mu_{SSJ}$ break the symmetry and are therefore expected to be small or at least technically natural small. The trilinear coupling $\mu_{SSJ}$ also break CP, as discussed in Sec.~\ref{sec:scatterings_scalars}.
For $m_S$ far above the electroweak scale, and a coupling hierarchy $\mu_{HHS}\gg \mu_{SSJ}$, the relevant processes will be $H\bar H\to SJ$ and $H S \to H J$. By solving numerically the Boltzmann equations \eqref{eq:eq_scat} for these processes, we find a very cold spectrum $\langle p/T\rangle_{\rm prod} \simeq 0.1$, in complete analogy to the toy model.
The correct relic density of $J$ is obtained for the couplings
\begin{align}
\frac{\mu_{HHS}\,\mu_{SSJ}}{\unit{TeV}\,\unit{keV}}\simeq 30\left(\frac{m_S}{\unit{TeV}}\right)^{5/2}
&&\text{ for }
&& m_J =\unit[7]{keV}\,.
\end{align}
We have thus managed to obtain very cold keV-scale DM by freeze-in scattering, simply by co-producing the light DM particle together with a much heavier particle that absorbs most of the scattering momentum. 

\begin{figure}[t]
\includegraphics[width=8cm]{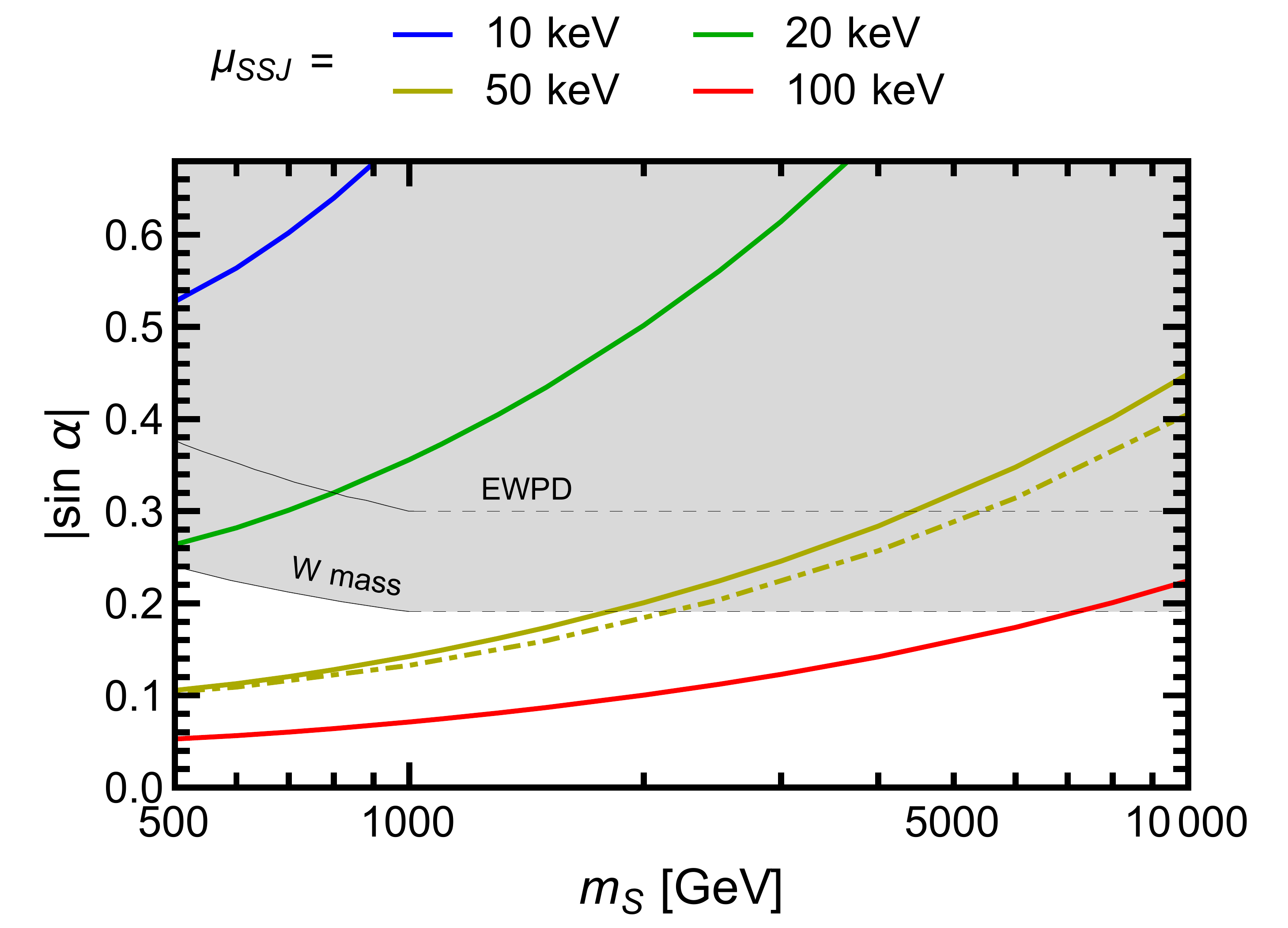}
\caption{The mixing angle $\alpha$, as given by \eqref{eq:alpha_heavy}, that gives the correct relic density for different values of $\mu_{SSJ}$. For the value $\mu_{SSJ} = \unit[50]{keV}$ we also plot the result obtained by neglecting the thermal mass of the Higgs doublet (dot-dashed line), setting it to $m_H = \unit[125]{GeV}$. The existing limits on the scalar mixing angle are taken from~\cite{Falkowski:2015iwa,Robens:2016xkb}.
The dotted parts denote the most conservative naive extrapolation of the given limits to larger values of $m_S$. Note that we have $\unit[7]{keV}$ DM with $\langle p/T \rangle_{\rm prod} \simeq 0.1$.
\label{fig:heavy}}
\end{figure}

Below the EWPT, $S$ and $H$ will acquire vacuum expectation values that also depend on the other couplings in the scalar potential that are irrelevant for the DM production. The unavoidable mixing of $S$ and the CP-even neutral scalar $h$ contained in $H$ thus becomes model dependent; however,  assuming that $\mu_{HHS}$ dominates the portal, we may obtain a mixing angle $\alpha$
\begin{align}\label{eq:alpha_heavy}
\sin\alpha\cos\alpha = \frac{\sin 2\alpha}{2} = \frac{\mu_{HHS} v}{m_S^2-m_H^2} \,.
\end{align}
In this case, we may trade the parameter $\mu_{HHS}$ for the mixing angle $\alpha$. The results for the value of $\alpha$ necessary to obtain the correct relic density are given in Fig.~\ref{fig:heavy} for different values of $\mu_{SSJ}$. There, we also show constraints on this mixing angle that arise from electroweak precision data (EWPD) and corrections to the $W$ mass~\cite{Falkowski:2015iwa,Robens:2016xkb}. Since the production temperature is above the EWPT, we have included the effect of the Higgs-doublet 1-loop  thermal mass (see e.g.~\cite{Cline:1993bd}), although the discrepancy with respect to using the $T=0$ Higgs-boson mass is found to be small, as also shown in the same figure.

From Fig.~\ref{fig:heavy} we see a numerical coincidence that makes this scenario (and the one below) particularly attractive: since the coupling $\mu_{SSJ}$ contributes to the mass of $J$ (via a self-energy diagram involving an $S$ loop) by an amount $\mathcal{O}(\mu_{SSJ})$, the requirement of not having large cancellations with the other possible sources of $m_J = \unit[7]{keV}$ fixes the scale of $\mu_{SSJ}$ to be at most in the tens of keV range. In turn, for $m_S \gg m_H$ one naturally has $\sin \alpha \lesssim m_H^2/m_S^2$ (barring again large cancellations), so that \emph{the correct relic density is obtained for $m_S \lesssim \mathcal{O}(\unit{TeV})$ and sizeable mixing angle}. Thus, in these scenarios the requirement of not having large cancellations implies that the scalar $S$ \emph{must} be in the phenomenologically interesting range, with observable mixing with $H$.

\begin{figure*}[t]
\includegraphics[height=5.2cm]{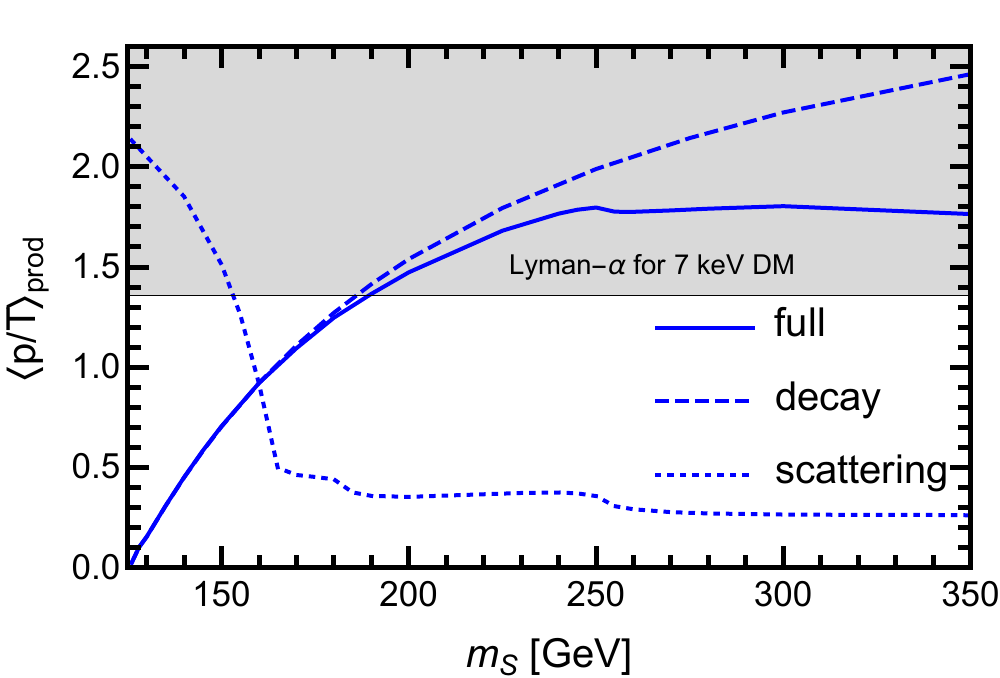}
\includegraphics[height=5.2cm]{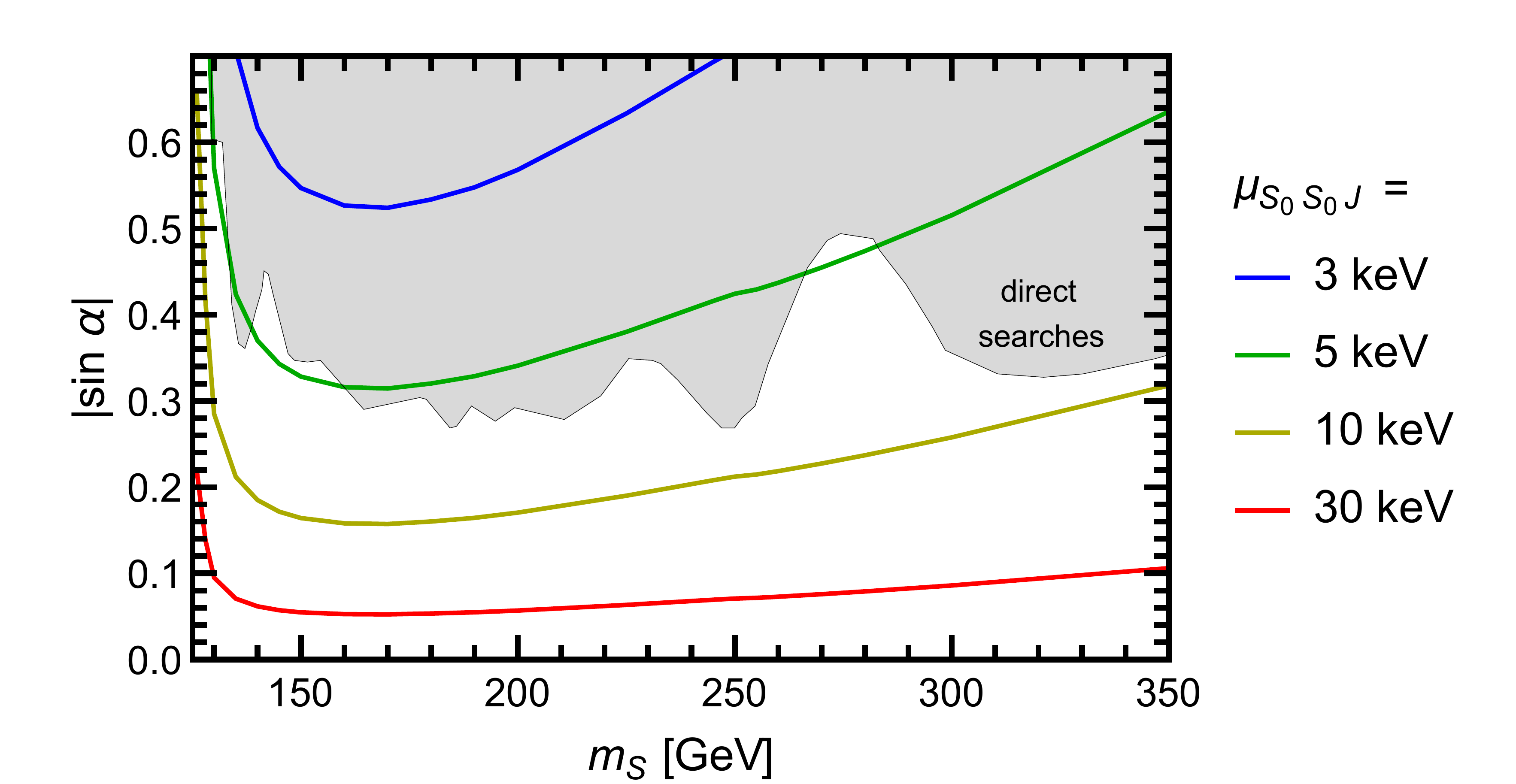}
\caption{
Average momentum of the dark matter (left panel) and value of the $S$--$h$ mixing angle $\alpha$ that gives the correct relic density (right panel) in Scenario D with $m_S > m_H$. The latter is given for different values of the coupling $\mu_{S_0 S_0 J}$. The existing limits on $\alpha$ (gray area) are taken from Ref.~\cite{Falkowski:2015iwa}. The gray region in the left panel is excluded by  the Lyman-$\alpha$ bound $m_{\rm TR} > \unit[4.65]{keV}$ when applied to a $\unit[7]{keV}$ DM mass, cf.~\eqref{eq:7kevLyman}.
}
\label{fig:medium}
\end{figure*}

\begin{figure*}[t]
\includegraphics[height=5.2cm]{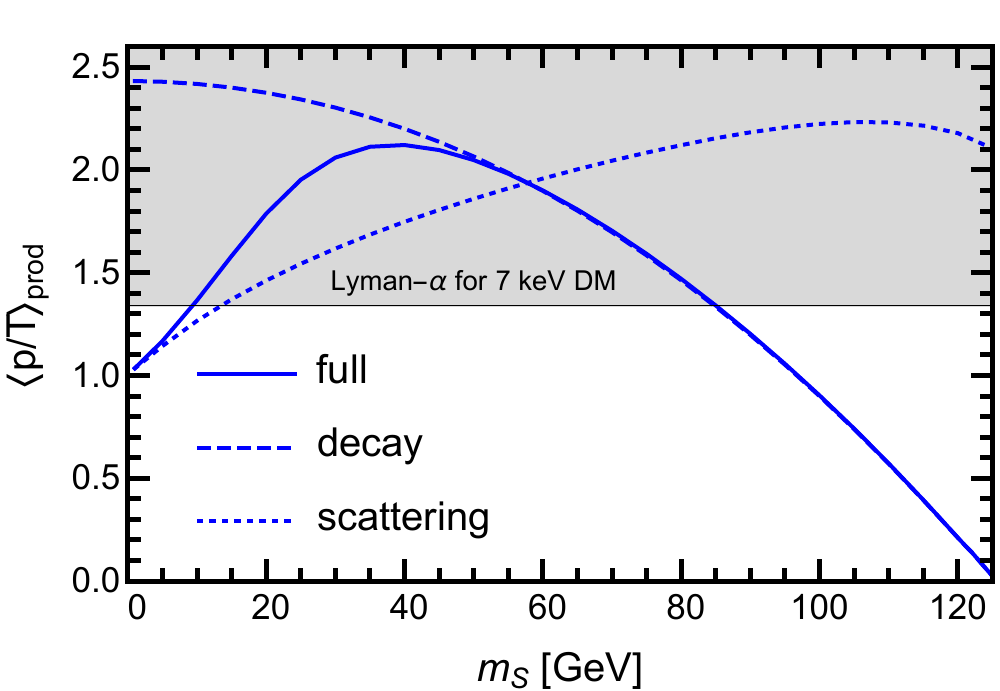}
\includegraphics[height=5.2cm]{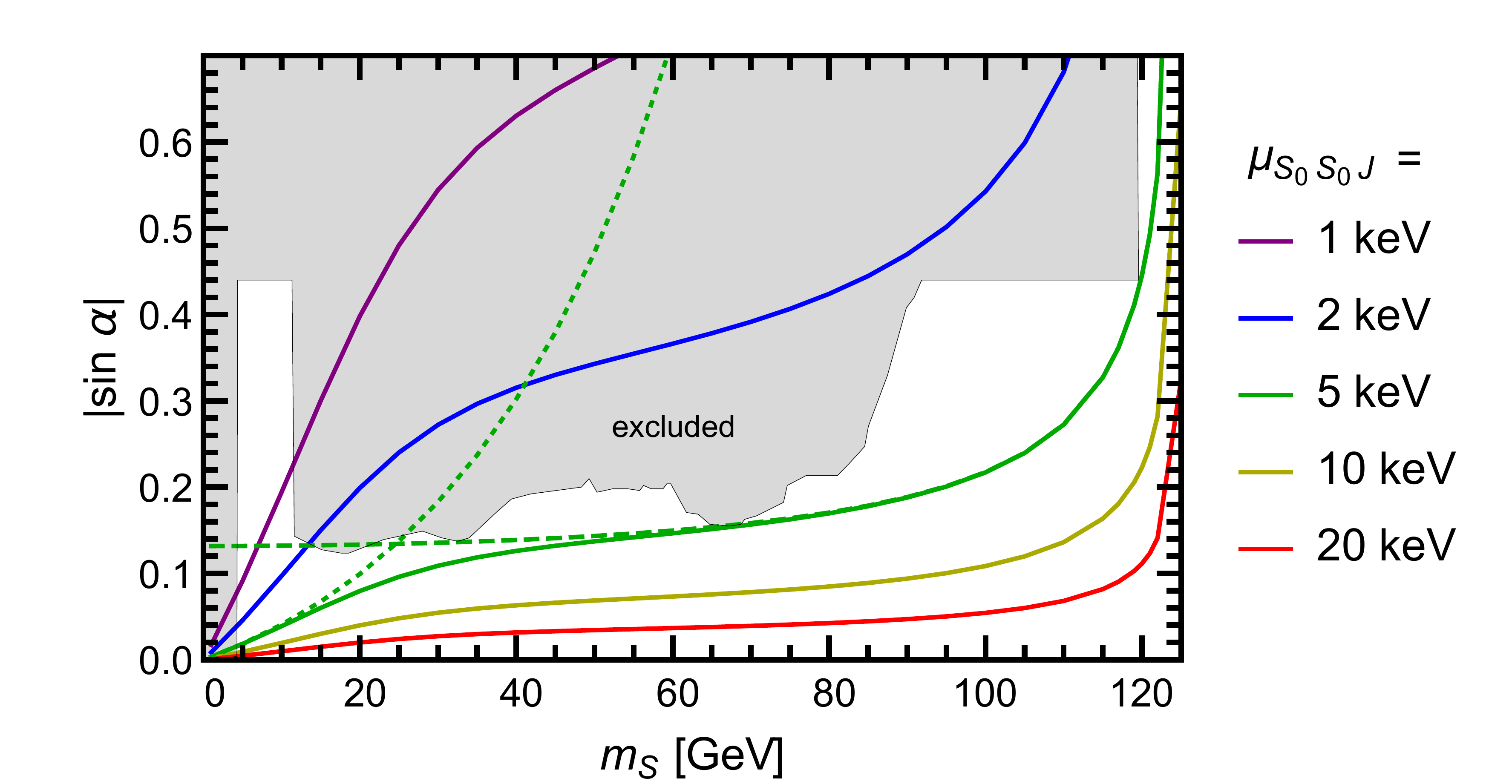}
\caption{The same as Fig.~\ref{fig:medium}, but for $m_S < m_H$. In the right panel, the exclusion limits come either from direct searches or Higgs-boson coupling measurements, the latter excluding $|\sin \alpha| \gtrsim 0.44$ for $m_S < \unit[120]{GeV}$~\cite{Falkowski:2015iwa}.
\label{fig:light}}
\end{figure*}

As for the quartic DM couplings not included in the Lagrangian~\eqref{eq:lagr_heavy}, they need to be $\lesssim \mathcal{O}(10^{-8})$ as already discussed in Section~\ref{sec:scatterings_scalars}. Notice that these small values do not imply the presence of fine tuning~\cite{Heurtier:2016iac}, since their radiative corrections are always significantly smaller than their tree-level value. For instance, the 1-loop contribution to the couplings $SS JJ$ and $H^\dag H JJ$ is of the order of $\mu_{SSJ}^2/(16 \pi^2 m_S^2) \sim 10^{-16}$. Finally, the potential interaction $H^\dag H J$ needs to have a coupling smaller than $\mu_{SSJ}$ in order to not contribute significantly to the production of $J$. Its role in the decay of $J$, that could generate the putative \unit[3.5]{keV} line, will be discussed in Sec.~\ref{sec:gamma_scalar}.

Lowering $m_S$ below \unit[500]{GeV} puts the DM production temperature near the EWPT, which severely complicates the analysis due to the dependence on the full scalar potential. We will not discuss this region in detail, whereas we pass to discuss the case in which the production is mostly below the EWPT.

\subsection{Light mediator scalar: decay and scattering}
\label{sec:light_scalar}

If DM is produced below the EWPT, a convenient parametrization for the required couplings is given by
\begin{align}
V &= \lambda_H \left(H^\dagger H - \frac{v^2}{2}\right)^2 + \frac12 \mu_S^2 S_0^2  + \frac12 m_J^2 J^2 \\
&\quad+ \mu_{HHS_0} \left(H^\dagger H - \frac{v^2}{2}\right) S_0 + \mu_{S_0S_0J} S_0^2 J\,,
\end{align}
where we neglected cubic and quartic couplings that are not of interest for the cold production of $J$. As above, the coefficients $m_J$ and $\mu_{S_0S_0J}$ break the $U(1)$ symmetry explicitly (the latter also breaking CP) and are hence expected to be small. We assume $\langle S_0\rangle = 0$, which in any case is just a redefinition of parameters for a real singlet scalar. In the unitary gauge, we have $H = (0,(h_0 + v)/\sqrt{2})^T$, and $h_0$ and $S_0$ mix into the mass eigenstates $h$ and $S$ with mixing angle $\alpha$ defined via
\begin{align}
\sin\alpha\cos\alpha = \frac{\sin 2\alpha}{2} = \frac{\mu_{HHS_0} v}{m_S^2-m_H^2} \,,
\end{align}
We will replace $\lambda_H$, $\mu_S$, and $\mu_{HHS_0}$ by the physical parameters $m_H$, $m_S$, and $\alpha$ in the following.
The mixing changes the cubic interactions of interest to
\begin{align}
V&\supset \mu_{SSJ} S^2 J + \mu_{hhJ} h^2 J+\mu_{ShJ}ShJ \\
&\quad+ \mu_{hSS} h S^2 + \mu_{hhS} h^2 S + \mu_{SSS} S^3+\mu_{hhh}h^3 ,
\end{align}
with coefficients
\begin{align}
\mu_{SSJ} &= \mu_{S_0S_0J} \cos^2\alpha\,,\\
\mu_{hhJ} &= \mu_{S_0S_0J} \sin^2\alpha\,,\\
\mu_{ShJ} &= -\mu_{S_0S_0J} \sin 2\alpha\,,\\
\mu_{hSS} &= \frac{\left(m_H^2+2 m_S^2\right) }{2 v}\cos \alpha \sin^2 \alpha\, ,\\
\mu_{hhS} &= \frac{\left(m_S^2+2 m_H^2\right) }{2 v}\cos^2 \alpha \sin \alpha\, ,\\
\mu_{SSS} &= \frac{3m_S^2}{2v}\sin^3 \alpha\,,\\
\mu_{hhh} &= \frac{3m_H^2}{2v}\cos^3 \alpha\,.
\end{align}
%The linear $J$ couplings could be further modified by a cubic term $\left(H^\dagger H - v^2/2\right) J$ which, as in Sec.~\ref{sec:heavy_scalar}, we assume small enough to not contribute to DM production.  

To leading order in $\alpha$, when $m_{S} > m_H$, DM is produced via the decay $S \to h J$ and the scattering processes $X S \to X J$, $X X \to S J$, with $X\in\{h,W,Z,t\}$. In Fig.~\ref{fig:medium} we show the results for $\langle p/T \rangle_{\rm prod}$ (left panel) and the mixing $\alpha$ that gives the correct relic density, the latter for various values of $\mu_{S_0 S_0 J} =  \unit[\mathcal{O}(10)]{keV}$. The $\unit[7]{keV}$ DM is sufficiently cold to satisfy the bound \eqref{eq:7kevLyman} for $\unit[125]{GeV} \lesssim m_S \lesssim \unit[190]{GeV}$. In this region, DM is mainly produced by decays, and its coldness stems from the partial phase-space closure of the decay $S \to h J$. Scattering would make DM cold for $m_S > 2 m_X$, but it turns out that the contribution of decays, which give hot DM for larger $m_S$, is rather dominant. For $m_S \gtrsim \unit[350]{GeV}$ DM is produced in part below and in part above the EWPT, and a more sophisticated analysis would be needed.

%\begin{figure}[t]
%\includegraphics[width=0.49\textwidth]{10}\\
%\includegraphics[width=0.49\textwidth]{80}
%\caption{Distribution function of DM, as function of the time variable $r = m_H/T$, for the model discussed in Sec.~\ref{sec:light_scalar} with $m_S = \unit[10]{GeV}$ (upper panel) or $m_S=\unit[80]{GeV}$ (lower panel). The dashed line denotes the rescaled thermal distribution, with an effective temperature given by $T_{\rm eff} = \langle p/T \rangle_{\rm prod}/\langle p/T \rangle_{\rm thermal}$. \label{fig:distr_fun}
%}
%\end{figure}

When $m_{S} < m_H$, the relevant decay process becomes $h \to S J$ and the results are plotted in Fig.~\ref{fig:light}. For relatively heavy $m_S$ decays still dominate, whereas for light $m_S$ scattering processes become dominant. Cold enough DM is obtained for $m_S \lesssim \unit[10]{GeV}$ or $\unit[85]{GeV} \lesssim m_S \lesssim \unit[125]{GeV}$, dominantly from scattering and decays, respectively. 
%In all cases in which we have obtained cold enough DM, the production is dominated by \emph{either} decays or scattering, so that DM distribution function approximately follows a rescaled thermal one (see Fig.~\ref{fig:distr_fun}) and an analysis in terms of $\langle p/T \rangle_{\rm prod}$, rather than on the full transfer function as discussed in Sec.~\ref{sec:multiple}, is sufficient for our purposes.
Finally, we point out that the discussion given in Sec.~\ref{sec:heavy_scalar} about the scale of $\mu_{S_0 S_0 J}$ and the other couplings, also apply to the regime discussed here.

\section{Radiative Decay -- the \texorpdfstring{$\unit[3.5]{keV}$}{3.5 keV} line}
\label{sec:decay}

In the previous sections we have presented several mechanisms to generate sufficiently cold keV-scale DM from in-equilibrium processes to satisfy Lyman-$\alpha$ constraints. The motivation comes in part from the tantalizing \unit[3.55]{keV} x-ray line observed in many different astrophysical objects, including Andromeda, Perseus, and the Galactic center~\cite{Bulbul:2014sua,Boyarsky:2014jta,Boyarsky:2014ska,Cappelluti:2017ywp}. Non-observation in stacked spectra of dwarf spheroidal satellites, on the other hand, put severe limits on the flux~\cite{Malyshev:2014xqa}.
Our aim is not to enter the discussion about the significance of this observation -- see Ref.~\cite{Adhikari:2016bei} for that -- but rather to take it as an interesting benchmark value. Several explanations of this line in terms of DM exist, arguably the simplest one being the two-body decay of either a \unit[7.1]{keV} DM fermion $F$ or boson $B$ with lifetime~\cite{Adhikari:2016bei}
\begin{align}
\tau (F\to\nu\gamma) &\simeq \unit[7\text{--}16\times 10^{27}]{s}\,,\\
\tau (B\to\gamma\gamma)&\simeq \unit[14\text{--}32\times 10^{27}]{s}\,.
\end{align}
As we have seen above, a DM mass of \unit[7]{keV} requires rather non-standard production mechanisms in order to satisfy Lyman-$\alpha$ constraints~\cite{Merle:2014xpa}.
Even if a viable production mechanism for DM is in place, for example the ones presented in the previous sections, the DM couplings to the SM necessary for the DM decay can lead to additional constraints on the model and need to be discussed.

We start our discussion with the well-known example of sterile-neutrino DM $N$.
In the simplest model, based on a tree-level coupling $y \overline{L}H N$, the relevant partial width can be calculated in terms of an active--sterile mixing angle~$\theta_\nu$,
\begin{align}
\Gamma (N\to \nu\gamma) &\simeq \frac{9\alpha}{2048\pi^4}\frac{m_N^5}{v^4} \sin^2 2\theta_\nu \\
	&\simeq \frac{1}{\unit[11\times 10^{27}]{s}} \left(\frac{m_N}{\unit[7.1]{keV}}\right)^5 \left(\frac{\theta_\nu }{3\times 10^{-6}}\right)^2. \nonumber
\end{align}
The mixing angle couples $N$ to the electroweak gauge bosons, but since the required angle is small, this does not lead to many other observable effects, so it is possible to accommodate the line in this way. One relevant effect is that a non-zero $\theta_\nu$ implies that a subcomponent of $N$ is unavoidably produced non-resonantly \`a la Dodelson--Widrow, i.e.~with a slightly hotter temperature, although this is typically not dangerous.
Since this kind of sterile-neutrino DM has been discussed at length in the literature~\cite{Adhikari:2016bei}, we will not go into any more details.

Focusing instead on bosonic DM $J$, we will distinguish three qualitatively different ways to obtain the decay into two photons. It is, of course, possible to have more than one mechanism operative.

\subsection{Mixing with the SM scalar}\label{sec:gamma_scalar}

If our  DM $J$ mixes with the SM scalar $h$ with an angle $\theta_h$ (thus violating CP if $J$ is a pseudoscalar), it will inherit all of its couplings, in particular the one-loop coupling $J F_{\mu\nu}F^{\mu\nu}$ that leads to diphoton decay, which for $m_J\ll m_e$ takes the simple form
	\begin{align}
	\Gamma (J\to\gamma\gamma) &\simeq \frac{121 \alpha^2}{2304\pi^3} \frac{m_J^3}{v^2} \sin^2\theta_h \label{eq:relaxion}\\
	&\simeq \frac{1}{\unit[14\times 10^{27}]{s}} \left(\frac{m_J}{\unit[7.1]{keV}}\right)^3 \left(\frac{\theta_h}{3\times 10^{-13}}\right)^2 . \nonumber
	\end{align}
	Note that the rate scales with $\theta^2 m_\mathrm{DM}^3/v^2$, so a much smaller mixing angle is required compared to the sterile neutrino case, where an extra suppression $m_\mathrm{DM}^2/v^2$ is present.
	At the Lagrangian level, the $J$--$h$ mixing can arise from the coupling $\mu_{hhJ} (H^\dagger H - v^2/2) \,J$ in the scalar potential, with $\mu_{hhJ} \simeq \unit[6\times 10^{-3}]{eV} (\theta_h/10^{-13})$. 
	The coupling to, e.g.~electrons, is then given by $J \bar{e} e\, \theta_h m_e/v $, which is sufficiently small to not lead to additional signals beyond the x-ray signal from $J\to\gamma\gamma$.
	A detailed discussion of the resulting phenomenology can be found in Ref.~\cite{Flacke:2016szy} in the context of relaxion models, where such a mixing with the Higgs is a crucial ingredient.
	In particular, the additional couplings are too small to thermalize $J$ in the early Universe or even freeze-in a relevant amount of $J$, in compatibility with the discussion in Sec.~\ref{sec:full_models}.
	
\subsection{Mixing with \texorpdfstring{$Z$}{Z}}

For a CP-odd scalar, an effective mixing $\theta_Z$ with the longitudinal component of the $Z$ boson, i.e.~the would-be Goldstone boson $G_Z$, can exist without violating CP. This generically occurs in majoron models and leads to a decay rate~\cite{Garcia-Cely:2017oco}
	\begin{align}
\Gamma (J\to\gamma\gamma) &\simeq \frac{\alpha^2 \theta_Z^2}{9216 \pi^3} \frac{m_J^7}{v^2 m_e^4} \label{eq:majoron}\\
	&\simeq \frac{1}{\unit[18\times 10^{27}]{s}} \left(\frac{m_J}{\unit[7.1]{keV}}\right)^7 \left(\frac{\theta_Z}{3\times 10^{-8}}\right)^2 . \nonumber
\end{align}
for $m_J \ll m_e$. 
Note that the amplitude for this process vanishes in the limit $m_J\to 0$ on account of anomaly freedom, which results in the additional suppression $m_\mathrm{DM}^4/m_e^4$ compared to the $J$--$h$ mixing, necessitating a larger mixing angle. 
In the singlet--triplet majoron model, the angle arises at tree level,
\begin{align}
\theta_Z \simeq \frac{2 v_T^2}{v f} 
\simeq 1.5\times 10^{-9}\left(\frac{v_T}{\unit[3]{GeV}}\right)^2 \left(\frac{\unit[5\times 10^7]{GeV}}{f}\right) ,
\end{align}
 $v_T$ ($f$) being the VEV of the triplet (singlet)~\cite{Schechter:1981cv,Bazzocchi:2008fh}; in the pure singlet-majoron case, it arises at loop level as 
\begin{align}
\theta_Z &\simeq \frac{\mathrm{tr}(m_D m_D^\dagger)}{16\pi^2 v f} \\
&\simeq 4\times 10^{-7} \left(\frac{\mathrm{tr}(m_D m_D^\dagger)}{4\pi v^2}\right) \left(\frac{\unit[5\times 10^7]{GeV}}{f}\right) ,\nonumber
\end{align}
$m_D$ being the standard Dirac mass matrix used in the seesaw mechanism. See Ref.~\cite{Garcia-Cely:2017oco} for the assumptions behind this result.
As shown in Ref.~\cite{Davidson:2001zk}, the matrix $m_D m_D^\dagger$ can be used to parametrize the high-energy part of the seesaw mechanism, and is in particular independent of the measured low-energy neutrino data. In our case it can be seen as a free parameter that determines the majoron coupling to charged fermions and photons, see Ref.~\cite{Garcia-Cely:2017oco} for more details.

One of the unique features of majoron models is the tree-level decay rate into active neutrinos,
\begin{align}
\Gamma (J\to\nu\nu)\simeq \frac{m_J}{16\pi f^2} \sum_\nu m_\nu^2\,.
\label{eq:majoron_to_neutrinos}
\end{align}
For sub-MeV majoron masses it would be incredibly difficult to directly detect such monochromatic neutrinos~\cite{Garcia-Cely:2017oco}, but the lifetime can still be constrained from cosmology to be above \unit[160]{Gyr} for cold DM~\cite{Audren:2014bca} (potentially up to \unit[170]{Gyr} depending on the dataset used~\cite{Poulin:2016nat}). This implies a lower bound on the $U(1)_L$ breaking scale of $f>\unit[5\times 10^7]{GeV}$ for a \unit[7]{keV} majoron, which holds for the normal neutrino mass hierarchy and becomes even stronger for inverted or quasi-degenerate spectra. Together with the upper bound on the triplet VEV of $v_T\lesssim\unit[3]{GeV}$ (at $3\sigma$) from the electroweak-precision parameter $\rho \simeq 1-2 v_T^2/v^2$~\cite{Olive:2016xmw}, this makes it difficult to have mixing angles $\theta_Z$ above $10^{-9}$ in the triplet--singlet majoron model. This differs from the conclusion in Refs.~\cite{Lattanzi:2013uza,Queiroz:2014yna}, in part because we use slightly stronger limits on $\Gamma (J\to\nu\nu)$. In the singlet model on the other hand, $\mathrm{tr}(m_D m_D^\dagger)$ can be as large as $4\pi v^2$ before reaching the non-perturbative regime, which allows $J$--$Z$ mixing angles as high as $\theta_Z\sim 4\times 10^{-7}$, an order of magnitude above the value required for the \unit[3.5]{keV} line. This could, of course, be considered as fine-tuning.

In all majoron models, the mixing $\theta_Z$ also induces couplings to, e.g.~electrons, given by $i J \bar{e} \gamma_5 e\, \theta_Z m_e/v $. Such couplings have been discussed in the context of axions, with typical limits from stellar cooling around $\theta_Z\sim 10^{-7}$~\cite{Raffelt:2012sp}, not dangerous for us.

This still leaves the question whether these small couplings could have an impact on the DM abundance via freeze in.
The strongest coupling here is to the heaviest particle, namely the top quark: $i J \bar{t} \gamma_5 t\, \theta_Z m_t/v $. Due to the large mass, this coupling is of order $10^{-8}$ in the region of interest, Eq.~\eqref{eq:majoron}. This could potentially play a role in cosmology, but a detailed discussion goes beyond the scope of this work.

\begin{table*}[t]
\renewcommand{\baselinestretch}{1.2}\normalsize
\centering
\begin{tabular}[t]{c|c|c|c}
\hline
\hline
 (main) process 				& $\langle p/T\rangle_\mathrm{prod}$ & Lyman-$\alpha$ for $m_\mathrm{DM}=\unit[7]{keV}$ & comments\\
 \hline 
$N_1 \to N_2 J$					&	$\mathcal{O}(m_J/m_1)$--$2.5$ 	& $0.7<m_2/m_1 <1$							&	family symmetry or inverse seesaw (Sec.~\ref{sec:RHNdecay_extended})\\
$H\overline{H}\to S J$	& $\mathcal{O}(0.1)$							& $\unit[500]{GeV} \lesssim m_S$	&	new scalar $S$ with $\alpha\sim 0.1$ (Sec.~\ref{sec:heavy_scalar})\\
$S\to h J$							& $\mathcal{O}(m_J/m_S)$--$2.5$		& $\unit[125]{GeV}\lesssim m_S\lesssim \unit[190]{GeV}$	&	new scalar $S$ with $\alpha\sim 0.1$ (Sec.~\ref{sec:light_scalar})\\
$h\to S J$							& $\mathcal{O}(m_J/m_h)$--$2.5$		& $\unit[85]{GeV}\lesssim m_S\lesssim \unit[125	]{GeV}$	&	new scalar $S$ with $\alpha\sim 0.1$ (Sec.~\ref{sec:light_scalar})\\
$X S\to X J \phantom{.}_{X=\{t,h,W,Z\}}$ 					& $1$--$2.5$											& $ m_S\lesssim \unit[10]{GeV}$	&	new scalar $S$ with $\alpha\sim 0.1$ (Sec.~\ref{sec:light_scalar})\\
\hline
\end{tabular}
\renewcommand{\baselinestretch}{1.0}\normalsize
\caption{\label{tab:working_models}
Models that can yield light cold DM $J$. The first column shows the main DM production process and the second the possible range of $\langle p/T\rangle_\mathrm{prod}$. In the third column we impose the Lyman-$\alpha$ bound from Eq.~\eqref{eq:translation2} for a $\unit[7]{keV}$ DM particle and derive the constraints on the heavy masses. In all cases, a decay $J\to\gamma\gamma$ can be induced by the mechanisms of Sec.~\ref{sec:decay}.}
\end{table*}

\subsection{Anomalies}
If $J$ is the pseudo-Goldstone boson of a $U(1)$ symmetry that is anomalous with respect to electromagnetism, a coupling to the pseudoscalar density $\tilde{F}_{\mu\nu} F^{\mu\nu}$ is generated,
\begin{align}
\L \ \supset \  \frac{\alpha E }{8\pi f} J \tilde{F}_{\mu\nu} F^{\mu\nu}  \,,
\end{align}
where $f$ is the $U(1)$ breaking scale, $\tilde{F}^{\mu\nu} \equiv \tfrac12 \epsilon^{\mu\nu\lambda\rho}F_{\lambda\rho}$ the dual field-strength tensor, and $E = \sum_X Q_X^2 Q_X^\mathrm{global}$ is the anomaly coefficient, summed over all chiral fermions $X$ with electric (global) charge $Q_X$ ($Q_X^\mathrm{global}$).
The decay rate is then given as~\cite{Olive:2016xmw}
\begin{align}
\Gamma (J\to\gamma\gamma) &= \frac{\alpha^2 E^2 }{256 \pi^3} \frac{m_J^3}{f^2} \label{eq:axion}\\
	&\simeq \frac{1}{\unit[25\times 10^{27}]{s}} \left(\frac{m_J}{\unit[7.1]{keV}}\right)^3 \left(\frac{\unit[3\times 10^{14}]{GeV}}{f/E}\right)^2 \! . \nonumber
\end{align}
 Axions are a prominent example here, with $E\neq 0$ typically induced when making the $U(1)$ anomalous with respect to color in order to solve the strong CP problem. 
In principle, arbitrary values for $E$ can be obtained by introducing particles beyond the SM, as long as they can either decay back into SM particles sufficiently fast to not modify cosmology or are too heavy to be produced at reheating~\cite{DiLuzio:2016sbl,DiLuzio:2017pfr}.
For the simplest invisible axion models, QCD leads to a relation of axion mass and breaking scale of the form $m_a \propto m_\pi f_\pi/f_a$, so $m_a\simeq\unit[7]{keV}$ would require a very low Peccei--Quinn breaking scale $f_a\simeq \unit[0.8]{TeV}$. This in turn leads to a large $K^-\to\pi^- a$ rate in conflict with experiments~\cite{Kim:1986ax}, even if we were tempted to fine-tune $E \simeq 10^{-12}$.

Since $f$ and $m_J$ are only directly linked in minimal axion models, we can simply pick a high scale $f$ and $E= \mathcal{O}(1)$ for our $\unit[7]{keV}$ DM~\cite{Higaki:2014zua,Jaeckel:2014qea}.
Note that this is the only decay mechanism so far that does not rely on DM--SM mixing; as such, the DM particle does not inherit any additional couplings beyond $J \tilde{F}_{\mu\nu} F^{\mu\nu}$, at least as long as we neglect the anomaly-inducing fermions and higher loop corrections.
A small amount of (rather hot) DM will unavoidably be produced thermally by the Primakoff process $\text{SM}\, \gamma\to \text{SM}\, J$, but will be suppressed if the reheating temperature is below the GUT scale~\cite{Jaeckel:2014qea}. 

\hspace{3ex}

Taking the $\unit[3.5]{keV}$ x-ray line seriously requires a DM decay rate that can be realized in several ways. As we have shown above, the required couplings or mixing angles are typically small enough to not produce too much DM. The DM decay can then often be separated from the DM production mechanism, for which we have provided new mechanisms that lead to cold enough light DM.

\section{Conclusion}
\label{sec:conclusion}

Dark matter with mass in the keV range is an interesting alternative to the standard WIMP scenario because it can lead to suppressed small-scale structures. The impact on structure formation depends however not simply on the DM mass, but rather its momentum distribution. 

In this article we have put forward several freeze-in production mechanisms that lead to rather cold keV-scale DM, essentially decoupling the DM mass from its average momentum.
In its simplest realization, one can obtain arbitrarily cold DM from the decay $A\to B\,\text{DM}$ if $m_A\sim m_B \gg m_\mathrm{DM}$, $A$ being in equilibrium with the SM. This merely requires the DM particle to have an off-diagonal coupling to two moderately degenerate heavy particles, easily found in many models.

A second class of processes that lead to cold light \emph{bosonic} DM can be identified by starting with an inverse decay $A\, B\to C$ with $m_A + m_B < m_C$ and emitting a DM particle from any of the particles involved. This requires a spin-diagonal coupling of the soft DM to one of the heavy particles $A$, $B$ or $C$ in order to obtain a resonant enhancement, which can in particular work for scalar or vector DM.

In Sec.~\ref{sec:full_models} we have discussed example models for a $\unit[7]{keV}$ bosonic DM candidate $J$ that satisfies the Lyman-$\alpha$ bounds by making use of these mechanisms. A summary of our findings is given in Table~\ref{tab:working_models}. Note that the degree of coldness is determined by kinematical features, so that the structure-formation constraints translate into ones for the mass spectrum of the involved particles. We stress again that the mechanisms discussed in Sec.~\ref{sec:toy_models} are more general than this and can be applied to other models.

All of these processes are of the freeze-in type, requiring a $\mathcal{O}(10^{-8})$ coupling of the DM particle, but the other new particles are in equilibrium with the SM. This is markedly different from the other production mechanism that is not in tension with Lyman-alpha data: the popular double freeze-in scenario for keV sterile-neutrino DM, where cold light DM is produced by the decay of a feebly interacting particle that is itself frozen in.
As a result, in our setup it is possible to search for the mediator particles to the light DM.

\section*{Acknowledgements}

We thank Laura Lopez Honorez for helpful discussions and comments on the manuscript.
JH is a postdoctoral researcher of the F.R.S.-FNRS; DT is supported by a ULB postdoctoral fellowship. 

%\vfill\null

\appendix

\section{Details on Boltzmann equations}
\label{sec:formulae}

Let us give further details on the derivation of the Boltzmann equations in Sec.~\ref{sec:technicalities}. In an expanding background, the Boltzmann equation for the DM distribution function $f(p,t) \equiv \sum_\mathrm{d.o.f.} f_\mathrm{DM}(|\mathbf{p}_\mathrm{DM}|,t)$, where the sum is over the internal degrees of freedom,  is
\begin{align}\label{eq:Boltz_master}
\bigg[\frac{\partial}{\partial t} - H p \frac{\partial}{\partial p}\bigg] f(p,t) =  \mathcal{C}(p) \,,
\end{align}
where $H = T^2/M_0$ is the Hubble parameter in the radiation-dominated epoch, with $M_0 \simeq M_\mathrm{Pl}/(1.66 \, g_*^{1/2})$. For a decay process $A \to B \, \mathrm{DM}$, the collision term is
\begin{align}
\begin{split}
\mathcal{C}(p)  &= \frac{1}{2 E_\mathrm{DM}} \int \!\! \frac{\dd^3 \mathbf{p}_{A}}{2 E_A (2\pi)^3} \int \!\! \frac{\dd^3 \mathbf{p}_{B}}{2 E_B (2\pi)^3} \\
& \quad\,\times (2 \pi)^4 \delta^{(4)}(p_A \!-\! p_B \!-\! p_\mathrm{DM}) \, |\mathcal{M}|^2 \, f_A(E_A) \,.
\end{split}
\label{eq:collis}
\end{align}
where $E_X \equiv E_X(\mathbf{p}_X)$ and $|\mathcal{M}|^2$ is the squared matrix element summed over initial and final degrees of freedom, following the conventions of Ref.~\cite{Luty:1992un}. Notice that the above expression, without additional factors of 2, is valid even if $B = \mathrm{DM}$. In writing down~\eqref{eq:collis} we have neglected quantum-statistics effects, which would instead give, in general, an additional dependence of $\mathcal{C}$ on $f_B$ and $f$. This dependence could be neglected, while retaining a quantum-statistics treatment, only if both $f_B, f \ll 1$, which would often not occur in the cases of interest in this work. Therefore, we adopt the classical statistics approximation systematically throughout the calculation.

Most of the integrals in Eq.~\eqref{eq:collis} can be performed either by symmetry or by conservation of 4-momentum, obtaining 
\begin{align}\label{eq:coll_explicit}
\mathcal{C}(p)  = \frac{|\mathcal{M}|^2}{16 \pi \, p^2} \int_{E^*}^{\infty} \!\! \dd E_A \, f_A(E_A)\,,
\end{align}
where we have exploited the fact that the invariant matrix element is a function of the masses and number of degrees of freedom only, and $E^*$ is the kinematical threshold for the process with a given $p$, i.e.
\begin{align}
E^* \equiv  \frac{m_A^2 - m_B^2}{4 \, p} \; +\; \frac{m_A^2}{m_A^2-m_B^2} \, p \,,
\end{align}
obtained by setting $m_\mathrm{DM} \to 0$. We can now switch to the variables $r \equiv m_H/T$ and $x \equiv p/T$. For bookkeeping purposes we have introduced the reference mass scale $m_H$, the Higgs-boson mass, which will of course drop out of physical quantities. Neglecting the change in $g_*$ during the time of production, one has:
\begin{equation}\label{eq:liouville}
\frac{\partial}{\partial t} - H p \frac{\partial}{\partial p} \ = \ \frac{m_H^2}{M_0 \, r} \, \frac{\partial}{\partial r} \;\;\bigg|_{x = \text{const.}} \;.
\end{equation}
Combining this with~\eqref{eq:Boltz_master} and~\eqref{eq:coll_explicit} one finally finds the Eq.~\eqref{eq:boltzmann_decay} we employ in the main text.

For the scattering process $A B \to C \, \mathrm{DM}$, the collision term is similarly given by
\begin{align}\label{eq:collis_scat}
&\mathcal{C}(p)  =  \frac{1}{2 E_{DM}} \int \!\! \frac{\dd^3 \mathbf{p}_{A}}{2 E_A (2\pi)^3} \int \!\! \frac{\dd^3 \mathbf{p}_{B}}{2 E_B (2\pi)^3} \int \!\!  \frac{\dd^3 \mathbf{p}_{C}}{2 E_C (2\pi)^3}  \notag\\
& \times (2 \pi)^4 \delta^{(4)}(p_A \!-\! p_B \!-\! p_C \!-\! p_\mathrm{DM}) \, |\mathcal{M}|^2 \, f_A(E_A) f_B(E_B)  \;.
\end{align}
Thanks to the classical-statistics approximation, we may rewrite this in a factorized form:
\begin{align}
&\mathcal{C}(p)  =  \frac{1}{2 E_{DM}} \int \!\! \frac{\dd^4 P}{(2\pi)^4} \frac{e^{-P_0/T}}{2 E_C} \, (2 \pi) \, \delta(E_C + E_\mathrm{DM} - P_0) \notag\\
& \int \!\! \frac{\dd^3 \mathbf{p}_{A}}{2 E_A (2\pi)^3} \int \!\! \frac{\dd^3 \mathbf{p}_{B}}{2 E_B (2\pi)^3}  \, (2 \pi)^4 \delta^{(4)}(p_A \!-\! p_B \!- \! P) \, |\mathcal{M}|^2 \;.
\end{align}
The second line is, up to a phase-space integral, the reduced cross section $\hat{\sigma}(s)$~\cite{Luty:1992un} for the process $C \, \mathrm{DM} \to A B$, with $s = P^2$, summed over initial and final degrees of freedom. Neglecting possible CP-violating effects, this is the same as~$\hat{\sigma}(s)$ for the original process $A B \to C \, \mathrm{DM}$. We thus find, setting also $m_\mathrm{DM} \to 0$,
\begin{align}
\mathcal{C}(p)  &= \frac{1}{16 \pi^2 p^2} \int_{s_{\rm min}}^{\infty} \!\! \dd s \; \frac{\hat{\sigma}(s)}{\left(1-\frac{m_C^2}{s} \right)} \,  \int_{P_{0}^*}^{\infty} \!\! \dd P_0 \, e^{- \frac{P_0}{T}}\,,
\end{align}
with the threshold energy (for a given $p$)
\begin{align}
P_0^* \equiv \bigg(1-\frac{m_C^2}{s} \bigg)^{-1} \!\! p \;+\; \frac{s - m_C^2}{4 \, p} \,,
\end{align}
and $s_{\rm min} =  \max\{(m_A+m_B)^2, m_C^2 \}$. Performing the $P_0$ integral, switching again to the variables $r$ and $x$ and combining with~\eqref{eq:Boltz_master} and~\eqref{eq:liouville}, we finally obtain our main-text Eq.~\eqref{eq:eq_scat}.

\section{Scattering cross sections}
\label{sec:xsecs}

In this appendix we give the scattering cross sections for the toy models of Sec.~\ref{sec:scatterings_scalars}, for simplicity in the limit $m_J = 0$. The quartic interaction of Eq.~\eqref{eq:quartic} gives the cross sections
\begin{align}
\sigma (S_1 S_1 \to S_2 J) &= \frac{\lambda _{\text{J112}}^2 \left(s-m_2^2\right)}{4 \pi  \sqrt{s^3 \left(s-4 m_1^2\right)}}\,,\\
\sigma (S_1 S_2 \to S_1 J) &= \frac{\lambda _{\text{J112}}^2 \left(1-m_1^2/s\right)}{4 \pi  \sqrt{\left(m_2^2-s\right){}^2+m_1^4-2 m_1^2 \left(m_2^2+s\right)}}\,.
\end{align}
For the cubic interactions of Eq.~\eqref{eq:cubic1} we assume $\mu_{\text{J11}}\ll\mu_{\text{112}}$ in order to neglect $S_i S_j \to JJ$ processes, leaving only $S_i S_j \to S_k J$ scatterings,
\begin{widetext}
\begin{align}
\sigma (S_1 S_1 \to S_2 J) &= \frac{2 \mu _{112}^2 \mu _{\text{J11}}^2 \left[\sqrt{s \left(s-4 m_1^2\right)}-2 m_1^2 \log \left(\frac{\sqrt{s}-\sqrt{s-4 m_1^2}}{\sqrt{s}+\sqrt{s-4 m_1^2}}\right)\right]}{\pi  m_1^2 s \left(s-4 m_1^2\right) \left(s-m_2^2\right)} \,,\\
\sigma (S_1 S_2 \to S_1 J) &= \frac{\mu _{112}^2 \mu _{\text{J11}}^2}{\pi  m_1^2 s \left(s-m_1^2\right) \left(\left(m_2^2-s\right){}^2+m_1^4-2 m_1^2 \left(m_2^2+s\right)\right)} \left[\left(m_1^2+s\right) \sqrt{\left(m_2^2-s\right){}^2+m_1^4-2 m_1^2 \left(m_2^2+s\right)}\right.\nonumber\\
&\quad\left.+2 m_1^2 s \log \left(\frac{m_1^2-m_2^2+s-\sqrt{\left(m_2^2-s\right){}^2+m_1^4-2 m_1^2 \left(m_2^2+s\right)}}{m_1^2-m_2^2+s+\sqrt{\left(m_2^2-s\right){}^2+m_1^4-2 m_1^2 \left(m_2^2+s\right)}}\right)\right] .
\end{align}
\end{widetext}
For the cubic interactions of Eq.~\eqref{eq:cubic2} we similarly assume $\mu_{\text{J22}}\ll\mu_{\text{112}}$, leaving us with the processes
\begin{align}
\sigma (S_1 S_1 \to S_2 J) &= \frac{\mu _{112}^2 \mu _{\text{J22}}^2}{\pi  \sqrt{s^3 \left(s-4 m_1^2\right)} \left(s-m_2^2\right)}\,,\\
\sigma (S_1 S_2 \to S_1 J) &= \frac{\mu _{112}^2 \mu _{\text{J22}}^2}{\pi  m_2^2 \left(s-m_1^2\right) }\\
&\quad \times \frac{1}{\sqrt{\left(m_2^2-s\right){}^2+m_1^4-2 m_1^2 \left(m_2^2+s\right)}}\,.\nonumber
\end{align}
The reduced cross sections relevant for the Boltzmann equations of interest can in all cases be obtained via
\begin{align}
\hat\sigma (S_i S_j\to S_k J ) = 2 \frac{\lambda (s,m_i^2,m_j^2)}{s}\sigma (S_i S_j\to S_k J )\,,
\label{eq:reducedxsec}
\end{align}
with the well-known K\"all\'en function $\lambda (a,b,c) \equiv a^2 +b^2+c^2-2 a b-2 a c - 2 b c$.
If the particles involved in the scattering carry spin or other internal degrees of freedom, they must be summed over in Eq.~\eqref{eq:reducedxsec}.

%\newpage

\bibliographystyle{utcaps_mod}
\bibliography{BIB}

\end{document}